\documentclass[preprint,showpacs,preprintnumbers,amsmath,amssymb,nofootinbib]{revtex4}

\usepackage{etex}

\usepackage{amsthm,amscd,amsbsy,array}
\usepackage{bm}% bold math
\usepackage{soul} % underline, strikethrough, etc.

\usepackage{graphics,graphicx,xcolor}

\usepackage{soul} % underline, strikethrough, 

\usepackage[colorlinks=true, pdfstartview=FitV, linkcolor=blue, citecolor=blue, urlcolor=blue]{hyperref} % hyperref

\newcommand{\red}{\textcolor{red}}  
\newcommand{\blue}{\textcolor{blue}}

\newcommand{\gb}{\colorbox{green}}

\newcommand{\dgreen}{\textcolor[rgb]{0,0.35,0}}

\newenvironment{redtext}{\color{red}}{\ignorespacesafterend} 
\newenvironment{bluetext}{\color{blue}}{\ignorespacesafterend}

\newcommand{\bblue}{\begin{bluetext}} 
\newcommand{\eblue}{\end{bluetext}} 
\newcommand{\bred}{\begin{redtext}}
\newcommand{\ered}{\end{redtext}}

\numberwithin{equation}{section}

\let\ssection=\section
\renewcommand{\section}{\setcounter{equation}{0}\ssection}

\newcommand{\bA}{{\bf A}}
\newcommand{\cA}{{\mathcal{A}}}

\newcommand{\bb}{{\bf b}}

\newcommand{\bone}{\boldsymbol{1}}

\newcommand{\bc}{{\mathbf{c}}}

\newcommand{\cC}{{\mathcal{C}}}

\newcommand{\diag}{\mathrm{diag}}

\newcommand{\da}{\dot{a}}

\newcommand{\db}{\dot{b}}

\newcommand{\dgamma}{\dot{\gamma}}
\newcommand{\dchi}{\dot{\chi}}
\newcommand{\ddchi}{\ddot{\chi}}

\newcommand{\e}{{e}} %%% Kinetic enrergy

\newcommand{\rg}{\mathrm{g}}

\newcommand{\cH}{{\mathcal{H}}}

\newcommand{\bk}{\mathbf{k}}

\newcommand{\bp}{{\bf p}}

\newcommand{\br}{{\bm{r}}}
\newcommand{\bx}{{\bm{x}}}

\newcommand{\bbR}{\mathbb{R}}

\newcommand{\Tr}{\mathrm{Tr}}

\newcommand{\bX}{{\bf X}}

\newcommand{\bY}{{\bf Y}}

\def\smallover#1/#2{\hbox{$\textstyle\frac{#1}{#2}$}} %
\def\Rarrow{\quad\Rightarrow\quad}

\def\bp{{\bm{p}}}

\def\benu{\begin{enumerate}}
\def\eenu{\end{enumerate}}
\def\beq{\begin{equation}}
\def\eeq{\end{equation}}
\def\beqa{\begin{eqnarray}}
\def\eeqa{\end{eqnarray}}
\def\nn{\nonumber}
\def\barray{\left(\begin{array}}
\def\earray{\end{array}\right)}
\def\barraynb{\begin{array}}
\def\earraynb{\end{array}}

\def\IS{S} % Round sphere
\def\?{\quad{\gb{\fbox{\texttt{?}}\;}}\quad}
\def\p{{\partial}}

\def\v0{\mathbf{0}}

\def\beq{\begin{equation}}
\def\eeq{\end{equation}}
\def\bea{\begin{eqnarray}}
\def\eea{\end{eqnarray}}

\def\p{\partial}

\def \p{{\partial}}

\def\6{\partial}
\def\7{\tilde}
\def\8{\widehat}
 \def\bx{{\bf x}}

\newcommand{\hbx}{{\hat{\mathbf{x}}}}
\newcommand{\hu}{{\hat{u}}}
\newcommand{\hv}{{\hat{v}}}

\newcommand{\const}{\mathop{\rm const.}\nolimits}
\newcommand{\half }{\frac{1}{2}}

\def\smallover#1/#2{\hbox{$\textstyle\frac{#1}{#2}$}} %
\def\smallcirc{{\raise 0.5pt \hbox{$\scriptstyle\circ$}}}
\def\2{{\smallover1/2}}

\let\ssection=\section
\renewcommand{\section}{\setcounter{equation}{0}\ssection}

\begin{document} % ££

\preprint{arXiv:1705.01378v4 [gr-qc]}

\title{Soft Gravitons \& the Memory Effect\\ for\\ Plane 
 Gravitational Waves
\\[6pt]}

\author{
P.-M. Zhang$^{1}$\footnote{e-mail:zhpm@impcas.ac.cn},
C. Duval$^{2}$\footnote{
%Aix-Marseille Universit\'e, CNRS, CPT, UMR 7332, 13288 Marseille, France.
%Universit\'e de Toulon, CNRS, CPT, UMR 7332, 83957 La Garde, France.
mailto:duval@cpt.univ-mrs.fr},
G. W. Gibbons$^{3,4,5}$\footnote{
mailto:G.W.Gibbons@damtp.cam.ac.uk},
P. A. Horvathy$^{1,4}$\footnote{mailto:horvathy@lmpt.univ-tours.fr},
}

\affiliation{
$^1$Institute of Modern Physics, Chinese Academy of Sciences, Lanzhou, China
\\
$^2$Aix Marseille Univ, Universit\'e de Toulon, CNRS, CPT, Marseille, France
\\
$^3$D.A.M.T.P., Cambridge University, U.K.
%\\ Wilberforce Road,
%\\ Cambridge CB3 0WA, U.K.},
\\
$^4$Laboratoire de Math\'ematiques et de Physique
Th\'eorique,
Universit\'e de Tours,
France
\\
$^5$LE STUDIUM, Loire Valley Institute for Advanced Studies, Tours and Orleans France
%%%%%%
%\texttt{arXiv: 1705.01378}
}

\baselineskip=16pt

\date{\today}


\begin{abstract} 
  The ``gravitational memory effect'' due to an exact plane wave provides us with an elementary description of the diffeomorphisms associated with the analogue of ``soft gravitons for this non-asymptotically flat system. We  explain how the presence of the latter may be detected by observing the motion of freely falling particles or other forms of gravitational wave detection.
    Numerical calculations confirm the relevance of the first, second and third time integrals of the Riemann tensor pointed out earlier.
    Solutions for various profiles are constructed.   It is also shown how to extend our treatment to Einstein-Maxwell plane waves and a midi-superspace quantization is given.
\\
% {\bf LongMemory PRD2 - final}
Phys.\ Rev.\ D {\bf 96} (2017) no.6,  064013\\
  doi:10.1103/PhysRevD.96.064013
\end{abstract}

\maketitle

\tableofcontents

%%%%%%%%%%%%%%%%%%%%%%%%%%%%%%%%%%%%%%%%%%%%%%%%%%%%%%%%%%%%%%%%%%%%%%%%%%%%%%
%%%%%%%%%%%%%%%%%%%%%%%%%%%%%%%%%%%%%%%%%%%%%%%%%%%%%%%%%%%%%%%%%%%%%%%%%%%%%%
\section{Introduction}\label{Intro}
%%%%%%%%%%%%%%%%%%%%%%%%%%%%%%%%%%%%%%%%%%%%%%%%%%%%%%%%%%%%%%%%%%%%%%%%%%%%%%
%%%%%%%%%%%%%%%%%%%%%%%%%%%%%%%%%%%%%%%%%%%%%%%%%%%%%%%%%%%%%%%%%%%%%%%%%%%%%%

The gravitational memory effect means, intuitively, that a  short burst of  gravitational wave  changes  the separation  of freely falling particles (viewed here as ``detectors'')  after the wave has passed \cite{ZelPol,BraGri}.
 The effect is potentially observable using LISA \cite{Fava}; after the first version of this paper was circulated, we were informed by P. Lasky that aLIGO might also be able to detect memory associated with binary black hole mergers in the not-too-distant future  \cite{Lasky:2016knh}. 
 The effect would be observed indirectly if the B-mode is detected in the CMB \cite{Kamionkowski:1996ks}. It is also relevant to recent work by Hawking, Perry,  and Strominger \cite{PeHaStPRL,PeHaSt1611} on soft graviton theorems in their attempt to resolve the ``Information Paradox"  of black hole physics.

Gravitational waves had long be thought to arise  from periodic sources such as binary star systems and were therefore expected to be detected through resonance. The novel idea of observing a burst-like gravitational wave through the displacement of freely falling bodies after the wave has passed was put forward in  1974 by Zel'dovich and Polnarev \cite{ZelPol} who suggested~:

\begin{quote}\textit{\narrower  \dots another, nonresonance, type of detector is possible, consisting of two noninteracting bodies (such as satellites). [\;\dots\;] 
  the distance between a pair of free bodies should change, and in principle this effect might possibly serve as a nonresonance detector. [\;\dots\;] One should note that although the distance between the free bodies will change, their relative velocity will actually become vanishingly small as the flyby event concludes.
}
\end{quote}

The idea of Zel'dovich and Polnarev was 
 elaborated  Braginsky and  Grish\-chuk \cite{BraGri},  who  introduced term  ``memory effect''. Both the title: \textit{ ``Kinematic resonance and the memory effect in
free mass gravitational antennas''} and the abstract of the latter paper give a clear idea of what is involved~:

\begin{quote}\textit{\narrower 
  Consideration is given to two effects in the motion of free masses
  subjected to gravitational waves, kinematic resonance and the memory effect.
  In kinematic resonance, a systematic variation in the distance between the
  free masses occurs, provided the masses are free in a suitable phase of the
  gravitational wave. In the memory effect, the distance between a pair of
  bodies is different from the initial distance in the presence of a  gravitational radiation pulse.
  Some possible applications [ \dots ]
  to detect gravitational radiation  \dots   }
\end{quote}  

Braginsky and Grishchuk were clearly concerned
with the motion of test masses (that is, no back reaction) moving in a weak gravitational wave.
Their analysis is at the \emph{linear} level.

 Two years later, Braginsky and Thorne \cite{BraTho} published a short Letter in \textit{Nature} making a distinction between two types of bursts, namely one \emph{without memory}, and one 
\emph{with memory}. 
The same distinction has been made earlier in \cite{Gibbons:1972}, but without the explicit introduction of the memory concept.
 
In the 1990s a \emph{nonlinear} form of memory  was discovered, independently, by Christodoulou \cite{Christo,Thor} and by Blanchet \& Damour \cite{BlaDam}.
 It arises from the contribution of the emitted gravitational waves  to the changing quadrupole and higher mass moments, cf. \cite{Gibbons:1972}. These  papers obtain  a permanent displacement.

Since the mid-1990s, there have been many studies of plane gravitational waves. As far as we are aware, few have dealt with the memory effect, and none with the concept of soft gravitons.
However after the first version of this paper appeared, our attention  was brought to two relevant papers of Harte,
\cite{Harte12,Harte15}. Although mainly concerned with optics, attention is drawn in \cite{Harte15} to a link with the memory effect that we shall elaborate later in this paper.

In what follows we consider the effect of a
fully non-linear plane gravitational wave on a detector whose back reaction is negligible.
This is done by considering geodesics in the exact plane wave background.
 We argue that our  approach, initiated in \cite{DGHZ17} following earlier work by Souriau \cite{Sou73}, is
 significantly simpler than those  in
\cite{BlaDam,Christo,Thor}, since it requires no knowledge of the source, nor
sophisticated understanding of non-linear partial differential equations.
Just simple calculation. It is based on the idea that
far from the source
we may approximate the gravitational wave in the
neighborhood of a detector
by an exact plane wave.

We shall  work in 3+1
spacetime dimensions  although
the  discussion of this paper readily generalises
to higher spatial dimensions. The assumption of three spatial
 dimensions is an obvious requirement for any discussion of
 physically realisable  detectors such as LIGO and LISA
 and moreover is also made
 in \cite{PeHaStPRL,PeHaSt1611}. However it has been pointed out
 that the boundary conditions for asymptotically flat
 higher dimensional spacetimes   differ considerably
 from those  in four spacetime dimensions \cite{Hollands:2016oma}
 which probably means that the obvious generalisation of the
 BMS group to higher dimensions \cite{Awada:1985by} is not applicable.
 
 A detailed  analysis  of  weak sources
 at the  linearised level analogous  to that in three-dimensions
 \cite{Gibbons:1972}
 indicates that there is no memory effect
 in higher dimensions \cite{Garfinkle:2017fre}. Since  the analysis
 of the present  paper reveals the importance of considering non-linear
 focussing effects  it may either be the case that these must
 be taken into account or our assumption that at large distance
 plane waves are a good approximation to outgoing gravitational 
waves fails in higher dimensions.

The plan of the paper is as follows. In section 
\ref{Planewaves}  we describe the basic geometry of plane gravitational waves and the two most useful coordinate systems used to describe them as well as the relation between them.
One referred to as Brinkmann (B) coordinates \cite{Bri} is global and allows the general vacuum solution
to be specified in terms of two arbitrary functions of a single
retarded  time variable. The second, called Baldwin-Jeffery-Rosen (BJR) coordinates \cite{BaJe,Ros} depends
upon the  same single retarded time variable and are adapted to 
a three-dimensional mutually commuting subset
of the five independent Killing vectors of plane wave spacetimes.
This fact  renders local  calculations simpler than in Brinkmann coordinates
for which only a single Killing vector is manifest.
The price to pay for this simplification is that the
metric is now specified by a $2\times 2$ symmetric matrix
giving the metric on the transverse space 
which requires solving a coupled system of Sturm-Liouville differential equations with no non-trivial global solution. This holds even in the locally flat case, as we show explicitly. 

Section \ref{Detectsec} is concerned with how gravitational waves are detected. This is at the heart of the gravitational memory effect and the detectability of soft gravitons.
We consider how a sandwich wave [i.e., one whose curvature
vanishes outside  a finite interval of retarded time] affects  freely falling particles initially at rest with respect to one another after the wave has passed.

In \ref{DetecTheory} and \ref{MemHamJac}  we recall how, in linear theory, this behaviour is encoded in
integrals of the Riemann
curvature with respect to retarded time and how these integrals serve as a diagnostic
for  the nature of the source. In sect. \ref{Cartography}   we show  how the memory
effect may be illustrated by means of ``Tissot'' diagrams illustrating the effect of gravitational pulses  on  a ring of freely falling particles.

Section \ref{GeoSec}  is concerned
with the detailed exact  behavior of these geodesics in the exact
plane wave backgrounds. In sects. \ref{BrinkmannGeo} and \ref{BJRGeoSec} we do this both in Brinkmann and
 in BJR coordinates. In (B)-coordinates our study is numerical however in the latter case we can proceed analytically~: by virtue
of Noether's theorem,  the spatial positions are independent of retarded
time. This has the consequence that for pulses, the memory effect is encoded  into a diffeomorphism (i.e. a coordinate transformation)
taking a part of  flat spacetime  in
standard inertial coordinates into  a patch of
flat space in non-inertial BJR coordinates. 
In  field theory approaches to general relativity,
such as those used in  \cite{PeHaStPRL,PeHaSt1611},
diffeomorphisms or coordinate transformations  are thought of as  
\emph{gravitational gauge transformations} and some gravitational
gauge transformations of  asymptotically flat spacetimes
are associated with \emph{soft gravitons}. In sec.
\ref{BefAfterGeo}. we argue that in our context, 
 flat plane waves in BJR coordinates correspond
to soft gravitons in the asymptotically flat spacetimes.

 In sec. \ref{PenroseSec}  we  relate our work
to the light cone structure of plane gravitational waves and a well-known analysis of Penrose. 

In sec. \ref{Einstein-MaxwellSec} we indicate how much of our work may be extended to  exact solutions of  Einstein-Maxwell theory. In particular, we point out that the coupled system has the Carroll symmetry identified recently \cite{DGHZ17} for pure gravitational waves.

Up to this point, our work has been purely classical.
In sec. \ref{MidiQSec} we turn to possible implications for the quantum  theory by considering a midi superspace (sec \ref{MidiSec}) made up of
plane gravitational  waves
and the associated space of quantum states.  

In the analogous case of electromagnetic waves there is an elaborate theory of polarization
and the photon states specified by Stoke's parameters
 correspond to points on what is called the  Poincar\'e
sphere which carries a Pancharatnam connection. In \ref{StokesSec}  we show how this formalism may be smoothly carried over to the case of gravitons.

The subject of plane gravitational waves has a long history
 and many contributions and reviews distributed over many different journals in many different languages. In reviewing the material necessary for an  exact and comprehensible  understanding of the memory effect and
 its relation to the concept of soft gravitons we have felt it necessary on the one hand to incorporate sufficient  material perhaps  well known to experts
 to make our account self-contained for non-experts while on the other hand giving sufficient credit to the pioneers of the field without overwhelming the reader  with an unmanageable list of all every contribution.
 
Some of the results presented here appear in summary in  \cite{ShortMemory}.

\goodbreak

%%%%%%%%%%%%%%%%%%%%%%%%%%%%%%%%%%%%%%%%%%%%%%%%%%%%%%%%
\section{Plane Gravitational Waves}\label{Planewaves}
%%%%%%%%%%%%%%%%%%%%%%%%%%%%%%%%%%%%%%%%%%%%%%%%%%%%%%%%

We begin by reviewing some facts about plane waves
\cite{Bri,Bondi,BoPiRo58,Peres59,Penr2,BaJe,Ros,ShortMemory,LaLi,Pirani:1956tn,Pirani:1956wr,EhlersKundt,exactsol,BoPi89, MiThoWh, Penr,GrifPod,vHolten97,Gibb75,Torre,BarHog,Shore,Sou73,Bini,Steinbauer}.

%%%%%%%%%%%%%%%%%%%%%%%%%%%%%%%%%%%%
\subsection{Brinkmann and Baldwin-Jeffery-Rosen coordinates}
%%%%%%%%%%%%%%%%%%%%%%%%%%%%%%%%%%%%

There are two commonly used coordinate systems for plane
gravitational waves, namely~:

\begin{itemize}

\item {\bf Brinkmann Coordinates} (B) \cite{Bri,Peres59} for which the metric is\footnote{Equation (\ref{Bplanewave}) gives the most general form of a pp-wave only in $D= 4$ total dimension; further components arise also if $D\geq5$ \cite{Bri}. In this paper, we limit ourselves to $D=4$.}
\beq
\rg =\delta_{ij}\,dX^idX^j+2dUdV+K_{ij}(U){X^i}{X^j}\,dU^2,
\label{Bplanewave}
\eeq
where the symmetric and traceless $2\times2$ matrix with components
$K_{ij}(U)$ characterizes the profile of the wave.
The  only  non-vanishing components of the Riemann tensor are, up to symmetry,\footnote{We use the convention $R^\mu_{\; \nu\rho\sigma}=2\partial_{[\rho}\Gamma^\mu_{\,\sigma]\nu}+\cdots$; indices are lowered according to $R_{\mu\nu\rho\sigma} = g_{\mu\lambda} R^\lambda_{\;\nu\rho\sigma}$.} 
\beq
R_{iUjU} (U)  = -  K_{ij}(U). 
\label{BRtensor}
\eeq
For suitable $K_{ij}$ the Brinkmann coordinates $(\bX,U,V)$, which are harmonic, are global \cite{Bondi,BoPiRo58}. The general form  of their profile is then 
\beq
K_{ij}(U){X^i}{X^j}=
\half{\cA_{+}}(U)\Big((X^1)^2-(X^2)^2\Big)+\cA_{\times}(U)\,X^1X^2,
\label{genBrink}
\eeq
where $\cA_{+}$ and $\cA_{\times}$ are the amplitude of the $+$ and $\times$ polarization state. 

\goodbreak
 Aside from their astrophysical applications to gravitational radiation, plane waves, in arbitrary spacetime dimensions, provide a general framework in which any ``natural'' non-relativistic dynamical system with a configuration space of dimension $n$ may be ``Eisenhart'' lifted to a system of null geodesics in an $(n+2)$ dimensional Lorentzian spacetime endowed with a covariantly constant null Killing vector field $\xi=\partial_V$
 \cite{Eisenhart,DBKP,DGH91,CaDuGiHo}.
 Conversely, a null reduction along the orbits of such  ``Bargmann'' spacetimes gives rise to a possibly time-dependent dynamical system on an $n$-dimensional configuration space.
From the ``Bargmann'' point of view, the metric (\ref{Bplanewave}) describes a non-relativistic particle subjected to an (attractive or repulsive)  
harmonic (and generally time-dependent and anisotropic) oscillator potential. 

\item{\bf Baldwin-Jeffery-Rosen Coordinates} (BJR) \cite{BaJe,Ros,LaLi}, for which
\beq
\rg=a_{ij}(u)\,dx^idx^j+2du\,dv,
\label{BJRmetrics}
\eeq
where the $2\times2$ matrix $a(u)=(a_{ij}(u))$ is strictly positive.
The BJR coordinates $(\bx,u,v)$ are not harmonic and are typically not global, but exhibit coordinate singularities \cite{Ros,Bondi,BoPiRo58,Pirani:1956wr,BoPi89,Sou73}, ---
a fact which gave rise to much  confusion in the early days of the subject.  Our investigations below provide further clarification of this issue.

\end{itemize} 

The relation between the two coordinates systems is given by
\cite{Gibb75,DGHZ17}
\beq
{\bX} =P(u)\,\bx,
\qquad
U=u,
\qquad 
V=v-\frac{1}{4}\bx\cdot\dot{a}(u)\bx,
\label{BBJRtrans}
\eeq
with  \footnote{The dot stands everywhere for the derivative w.r.t. $u$.},
\beq
a(u)=P(u)^{T}{}P(u),
\label{aPP}
\eeq
where $P$ satisfies
\beq
\ddot{P}=K\,P.
\label{SLB}
\eeq
For a given matrix $K$, this is a second-order ODE of the Sturm-Liouville type for $P$, which implies that $P^{T}\dot{P}-\dot{P^{T}}P=\const$ Then the initial values of $\dot{P}$ and of $P$ may be chosen so that the constant vanishes,
\beq
P^{T}\dot{P}-\dot{P^{T}}P=0.
\label{2ndcond}
\eeq
The mapping (\ref{BBJRtrans}) transforms the quadratic ``potential''  $K_{ij}(U){X^i}{X^j}$ in (\ref{Bplanewave}) into a time-dependent transverse metric (\ref{aPP}) and vice versa. The relation  is
\beq
K= \half P \bigl(\dot{b} + \half b^2\bigr) P^{-1},
\qquad
b= a^{-1} \dot a.  
\label{Einstein}
\eeq

%%%%%%%%%%%%%%%%%%%%%%%%%%%%%%%%%%%%%%%%%%
\subsection{Plane waves in BJR coordinates}\label{PlaneWavesBJRsection}
%%%%%%%%%%%%%%%%%%%%%%%%%%%%%%%%%%%%%%%%%%

Up to symmetry,
the only non-zero components of the Riemann tensor are, \begin{equation}
R_{uiuj}=-\half \Big(\ddot{a}-\half\dot{a}a^{-1}\dot{a}\Big)_{ij}\,,
\label{RBJR}
\end{equation} 
yielding the Ricci tensor, whose  only nonzero component is (\ref{Einstein}),
\begin{equation}
R_{uu}=-\half\Tr\left(\dot{b}+\half{}b^2\right)
\quad\text{with}\quad b=a^{-1}\dot{a}.
\label{RicciBis}
\end{equation}
The most general flat metric obtained by solving the equation
$R_{uiuj}=0$. With  initial conditions  
\beq
a_0=a(u_0)
\qquad\text{and}\qquad
\da_0=\da(u_0)
\label{initacond}
\eeq 
we find 
\begin{eqnarray}
a(u)=\left(a_0+\half(u-u_0)\da_0\right)a_0^{-1}\left(a_0+\half(u-u_0)\da_0\right),
\label{II13}
\end{eqnarray}
from which we infer that 
\begin{equation}
a(u)=a_0^\half\Big(\bone+(u-u_0)c_0\Big)^2a_0^\half
\qquad
\text{where}
\qquad
c_0=\half{}a_0^{-\half}\da_0\,a_0^{-\half}
\label{aflat}
\end{equation}
where $a_0^\half$ is a (symmetric) square-root of the positive matrix $a_0$.

If, in particular, the initial conditions in (\ref{initacond})  are 
$ 
a_0=\bone
\,\text{\small and}\,
\da_0=0,
$
then we obtain flat spacetime in inertial coordinates, for which
$a(u)=\bone$ for all $u$. More generally,  (\ref{aflat})  allows us to recast, in any flat  region, the metric (\ref{BJRmetrics}) into standard Minkowskian form by a change of coordinates,
$(\bx,u,v)\mapsto (\hbx,\hu, \hv)$. For
\begin{subequations}
\begin{align}
\label{xhx}
\hbx&=\left(\bone+(u-u_0)c_0\right)a_0^\half\bx,
\\
\label{uhu}
\hu&= u,
\\
\label{vhv}
\hv&=v-\half\bx\cdot{}\left(a_0^\half{}c_0\left(\bone+(u-u_0)c_0\right)a_0^\half{}\bx\right),
\end{align}
\label{NewCoordinates}
\end{subequations}
\vskip-7mm
whose inverse is
\vskip-8mm
\begin{subequations}
\begin{align}
\label{hxx}
\bx&=a_0^{-\half}\left(\bone+(u-u_0)c_0\right)^{-1}\hbx,
\\
u&=\hu,
\\
v&=\hv+\half\hbx\cdot{}\left({}c_0\left(\bone+(u-u_0)c_0\right)^{-1}\hbx\right),
\end{align}
\label{NewCoordinatesInv}
\end{subequations}
\vskip-5mm
one readily finds indeed that
\begin{eqnarray}
\rg=d\bx\cdot a(u)d\bx+2du\,dv
=d\hbx\cdot d\hbx+2d\hu\,d\hv.
\end{eqnarray}
We will call $(\hbx,\hu,\hv)$  a \textit{manifestly flat BJR coordinate chart}.  

Two metrics related by a coordinate transformation,
i.e., by a diffeomorphism, are usually regarded as equivalent.

However as it stands, this statement is not very precise.
One needs to specify how the diffeomorphism $f$ 
acts on the spacetime $\{ {\cal M}, g\}$ under consideration. If it is the identity outside a compact  set within the spacetime manifold ${\cal M}$ (which we assume to be non-compact), then one typically assumes
then the two spacetimes $\{ {\cal M}, g \}$ and $\{{\cal M}, f^\star g \}$, where $f^\star$ denotes pull back, are physically equivalent,
  i.e. ``mere coordinate transformations of one another''.
  
  However if the diffeomorphism $f$  does not vanish outside a compact set and does not tend in some appropriate sense to the identity at ``infinity'', more care is required.
  For example in (suitably defined) asymptotically flat spacetimes, there  is a class of of distinguished coordinate systems related by the subset of diffeomorphisms which do not tend to the identity at infinity, but which nevertheless
 take asymptotically flat spacetimes to asymptotically flat spacetimes
  \footnote{One should beware
   that confusion can arise if the class of permissible diffeomorphisms are not viewed \emph{actively}. 
   In discussing  the  elementary geometry of Euclidean space,
   the  metric tensor in Cartesian coordinates  and  in spherical coordinates differ substantially,  as do the coordinate  functions themselves at  large radius. However viewed {\sl passively}  a ``change of coordinates'' merely relabels the points which are left fixed.}.
  The set of such diffeomorphisms is referred to as the \emph{asymptotic symmetry group. }
  These include the translations and boosts.
  
Solutions of the Einstein equations which differ by such
  diffeomorphisms are not usually thought of as physically identical since they could, for example, describe two black holes moving towards one another. 
  
In $3+1$ spacetime dimensions the asymptotic symmetry
  is well known to be the infinite dimensional BMS group. Spacetimes
  which differ by the action of elements  of the BMS group are typically
  regarded as physically distinct. This is especially so in
  scattering theory, both at the  classical level 
  and in attempts to construct a perturbative
  quantum version in which the S-matrix plays
  an important  role, the classical theory being described by
  tree-diagrams. \footnote{One might object
    that strictly from a   rigorous point of view,
    no  S-matrix exists in quantum fields
    theories based on standard  Fock space
    constructions of their Hilbert spaces
    \cite{Haag:1955ev}  and even  classically the soundness
    so-called Lorentz Covariant  approaches
    has often  been questioned on causality grounds
    (see e.g. \cite{Penrose:1980})
    but in this paper we shall set aside such doubts.}
    
Since in this approach  gravitational waves carrying arbitrarily small energy have to be considered, the quantum theory has to address
  certain difficulties, specifically infinite quantities which arise even in electromagnetic theory in
  Minkowski spacetime where they are ascribed
  to the presence of  so-called soft (i.e. zero energy)  photons.
  At the quantum level these soft quanta are frequently
  assigned states in the quantum Hilbert space. At the classical level
  these soft photons  carry  vanishing electromagnetic  fields
  and so differ only by  electromagnetic gauge transformations which however do
  not tend to the identity at infinity, ``at infinity''  being, in this
  case, a neighbourhood of the conformal boundary of Minkowski spacetime
  \footnote{We are grateful to Piotr Bizon \cite{Bizon} for informing us
    of what appears to the first mention of a memory effect in the
    electromagnetic case \cite{Staruszkiewicz:1981mq}. We subsequently learnt
    from Malcolm Perry \cite{Perry} that an even earlier though
   not very explicit mention may  well come from Mott in a paper in which
     he computed the number of photons produced in Rutherford scattering \cite{Mott}.}.
  The work of \cite{PeHaStPRL,PeHaSt1611} is an
  attempt to make use of  much earlier work by themselves and others (referred to in detail in their papers ),  which
  extends the ideas and results  obtained for photons to gravitons.

The starting point  of \cite{PeHaStPRL,PeHaSt1611}
 (which has not been without its critics
 \cite{Mirbabayi:2016axw,Bousso:2017dny,Bousso:2017rsx}),
 was to consider asymptotically flat spacetimes and the BMS group. The idea of the present paper
 is to consider a much simpler situation: plane wave spacetimes.
 In the present case  the diffeomorphism defined by
 (\ref{NewCoordinates})--(\ref{NewCoordinatesInv})
 does not tend to identity  as $|\bx|$ or $|u|$ tend to infinity.
 Moreover, since every metric tensor given by (\ref{aflat}) is
locally flat, it is tempting  to regard them,
in the language of quantum field theory, as ground states or vacua.  

As we recalled above, in theories with
  no massless excitations one usually regards all such ``gauge equivalent'' vacua as equivalent. But in theories with massless excitations 
 it is customary to regard such vacua related by gauge transformations which do not tend to identity at infinity
as \emph{non-equivalent}, differing by the presence of  ``soft'' (i.e., zero-energy) quanta. Such claims are often supported by a canonical or Hamiltonian
treatment in which the soft states are associated with charges or moment maps which may be expressed as surfaces integrals ``at  infinity''
which, for asymptotically flat spacetimes, are 2-surface integrals evaluated
on the conformal boundary. This has been done in the asymptotically flat case in \cite{PeHaStPRL,PeHaSt1611,Compere:2016jwb}. 

In the present case the massless excitations correspond,
at the quantum level, to gravitons and so one may
regard the metrics given by (\ref{aflat})
as ``dressed by  soft gravitons'',  (i.e. carrying
vanishing energy) the dressing being affected by
the pulse of gravitational radiation itself made up of ``hard''
(i.e. carrying non-vanishing energy)
gravitons.
This interpretation is consistent with that given in \cite{PeHaStPRL,PeHaSt1611} in the asymptotically flat case. 

To confirm this  suggestion in full mathematical detail would require a detailed treatment of what one means by ``at infinity'' for plane gravitational waves, their conformal
 boundary (c.f. \cite{Marolf:2002ye}), a canonical or Hamiltonian treatment, the
 identification of possible moment maps defined
 as 2-surface integrals ``at infinity''.
 This is an interesting and demanding  challenge for the future. For the present we shall content ourselves with fleshing out some
 aspects  of  plane gravitational waves  which (we feel) make
 our suggestion plausible at the physical level.
 As partial compensation we note that being based
 on exact solutions of the Einstein equations our results evade the strictures  of \cite{Penrose:1980} alluded to earlier.
     
We will consider sandwich waves i.e. gravitational waves  which are
flat \textit{outside} the sandwich but not \textit{inside}, i.e. for
 $u\in[u_i,u_f]$.
Our point here is that flat spacetimes  both in the ``before-zone'' 
$u<u_i$ and in the ``after-zone'' 
$u>u_f$ \cite{BoPi89} are  \emph{non-equivalent}. 

Inside a sandwich wave we only have Ricci-flatness cf.  (\ref{RicciBis}), 
\begin{equation}
\Tr\left(\db+\half{}b^2\right)=0.
\label{Ricci0BJR}
\end{equation}
By (\ref{Einstein}) this is precisely the tracelessness of $K$. 

BJR coordinates  are convenient for comparing the standard linear theory in transverse traceless gauge with the fully non-linear theory. For plane waves in linear theory one has a metric of the form (\ref{BJRmetrics}) with
\beq
a_{ij}= \delta_{ij} + h_{ij}(u) +\dots
\eeq
Thus 
\beq
P_{ij}(u)= \delta_{ij} + \half\,h_{ij}(u) +\dots\;,
\qquad
K_{ij}(u)=  \half\,\ddot{h}_{ij}(u) +\dots
\eeq
Thus after the wave has passed, i.e. if $K_{ij}=0$, we have 
$h_{ij}(u)=h^{0}_{ij}+u\,h^{1}_{ij}$  where $h^{0}_{ij}$ and $h^{1}_{ij}$ are independent of $u$. If $h^{1}_{ij}=0$ we have the metric \# (5.19) of Favata \cite{Favata08} in his discussion of the possibilities of detecting the memory effect with interferometers and his \#(5.20) transforming to manifestly flat coordinates. These agree with 
(\ref{II13})  and (\ref{NewCoordinates}). Note that generically $h_{ij}$ is \emph{linear} in $u$.

To see that the BJR coordinates are indeed necessarily singular as stated,  let us
define 
\begin{equation}
\chi=\big(\det{a}\big)^{\frac{1}{4}}>0
\qquad
\&
\qquad
\gamma=\chi^{-2}a,
\label{chi-gamma}
\end{equation}  
so that $b=\gamma^{-1}\dgamma+2\chi^{-1}\dchi\,\bone$. 
Since $\det\gamma=1$, we readily obtain $\Tr(\gamma^{-1}\dgamma)=0$; this allows us to show that  
(\ref{Ricci0BJR}) is equivalent to the Sturm-Liouville equation
\begin{equation}
\ddchi+\omega^2(u)\chi=0,
\qquad
\omega^2(u)=\frac{1}{8}\Tr\left((\gamma^{-1}\dgamma)^2\right)
\label{oscieqn}
\end{equation}
which thus guarantees that the vacuum Einstein equations are satisfied for an otherwise arbitrary choice of the unimodular symmetric $2\times2$ matrix, 
\begin{equation}
\gamma(u)
=
\left(
\begin{array}{lr}
\alpha(u) & \beta(u)\\
\beta(u) & (1+\beta(u)^2)/\alpha(u)
\end{array}
\right).
\label{gamma}
\end{equation}
Thus the matrix $a(u)$ depends  on two arbitrary functions $\alpha(u)$ and $\beta(u)$, see Eqn (\ref{gamma}) and \cite{DGHZ17}.

The  positivity of the matrix $(\gamma^{-1}\dgamma)^2$ 
implies that $\omega^2$ in (\ref{oscieqn}) is positive;  
the equation describes therefore an attractive oscillator with a
time-dependent frequency. It follows that $\chi(u)$ is a concave function, $\ddot{\chi}<0$, which in turn implies the vanishing of $\chi$ for some $ u_\mathrm{sing} > u_i$,
\beq
\chi(u_\mathrm{sing})=0,
\label{chizero}
\eeq
 signalling a singularity of the metric (\ref{BJRmetrics}).
Choosing $
a(u)=\diag(a_{11},a_{22}),
$ 
for example, we find,
\beq
\omega^2(u)=\frac{1}{16}\left(
\frac{\dot{a}_{11}}{a_{11}}-\frac{\dot{a}_{22}}{a_{22}}\right)^2
\label{ofreq}
\eeq
and the Sturm-Liouville equation (\ref{oscieqn}) becomes
\beq
\dfrac {\ddot a_{11}} {a_{11}} + \dfrac {\ddot a_{22}} {a_{22}} 
- \dfrac {1}{2} (\dfrac { \dot a_{11} ^2 } {a_{11}^2 } +
\dfrac { \dot a_{22}^2} {a_{22}^2} ) = 0.
\label{adiagSL}
\eeq
Expressed in terms of  the matrix $P$, this is simply,
\beq
\dfrac {\ddot P_{11}}{P_{11}} + \dfrac {\ddot P_{22}} {P_{22}} = 0
\label{PdiagSL}
\eeq
which is indeed  $\Tr\, K = 0$ since $\ddot{P}P^{-1}=K$.
%%%%%%%%%%%%%%%%%%
\begin{figure}[h]
\includegraphics[scale=.2]{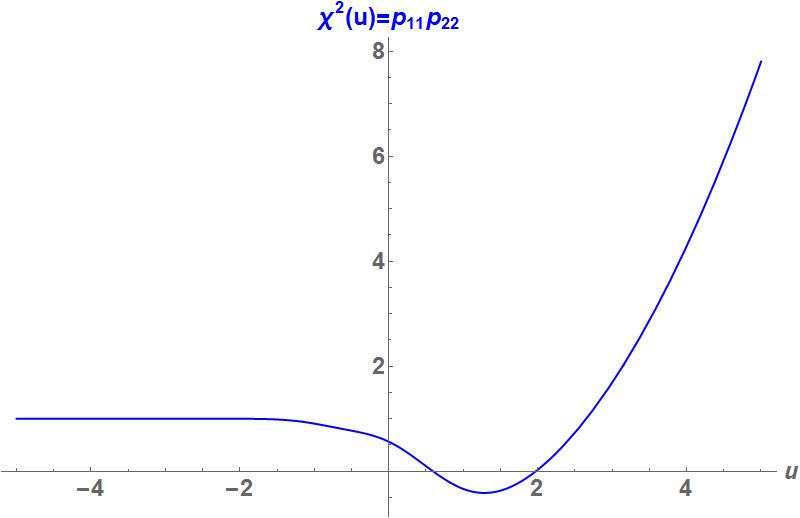}\;\;\;
\includegraphics[scale=.21]{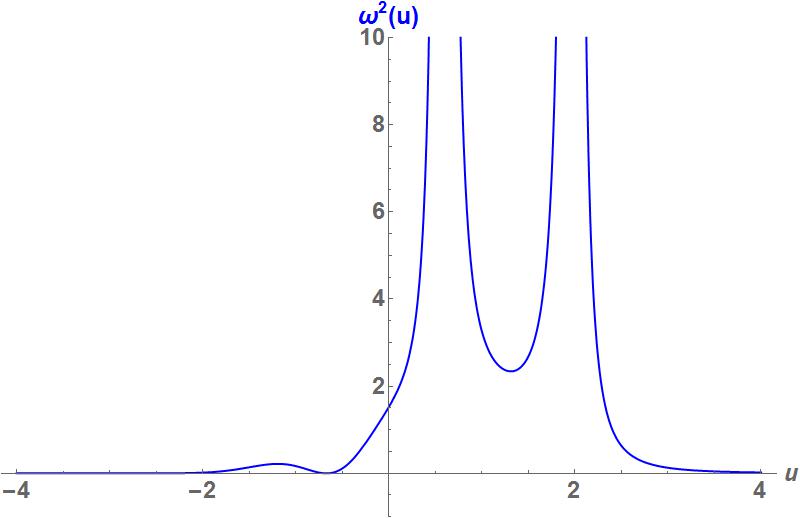}\\
 (a) \hskip 64mm (b)\\
\vskip-3mm 
\caption{\textit{(a) $\chi=(\det{a})^{1/4}$ and (b) $\omega^2$ in (\ref{oscieqn}), respectively, calculated numerically for $\cA_+=\cA^3_+$
 confirm that, in BJR coordinates, the metric becomes singular at $u_\mathrm{sing}$. }}
\label{chifig}
\end{figure}
%%%%%%%%%%%%%%%%%

While we  can not solve the non-linear equation
(\ref{PdiagSL}) in general,
we may proceed differently~: starting with some physically relevant profile in Brinkmann coordinates and 
solving (\ref{SLB}) numerically allows us to calculate the matrix $a$ and to plot
 $\chi(u)$ and $\omega^2(u)$ in (\ref{chi-gamma}) and (\ref{oscieqn}). This confirms 
the existence of a $u_\mathrm{sing}>u_i$ such that the metric becomes singular, $\chi(u_\mathrm{sing})=0$. 
For the choice $\cA_{\times}=0\,,\,\cA_+=\cA^3_+$  in (\ref{Kd3}) [justified in the next section], for example,
$\chi$ and $\omega^2$  are  plotted in Fig.\ref{chifig}. 

\goodbreak
%%%%%%%%%%%%%%%%%%%%%%%%%%%%%%%%%%%%%%
%%%%%%%%%%%%%%%%%%%%%%%%%%%%%%%%%%%%%%
\section{Detection of the Memory Effect}\label{Detectsec}
%%%%%%%%%%%%%%%%%%%%%%%%%%%%%%%%%%%%%%
%%%%%%%%%%%%%%%%%%%%%%%%%%%%%%%%%%%%%%

%%%%%%%%%%%%%%%%%%%%%%%%%%%%%%%%%%%%%%%
\subsection{Detection theory}\label{DetecTheory}
%%%%%%%%%%%%%%%%%%%%%%%%%%%%%%%%%%%%%%%

We turn now to the question of the detectability of soft gravitons. As pointed out in  the pioneering papers of Pirani
\cite{Pirani:1956tn,Pirani:1956wr} knowledge of the 
\emph{relative}  motion of freely falling
particles  in time-dependent gravitational fields
is essential  for our understanding of gravitational radiation
and its detection. In practical devices the ``particles'', such as
mirrors in interferometers, or the individual atoms
in old fashioned bar detectors   are
never truly freely falling since they are subject to
various forces holding them in place. Nevertheless it is the
relative motions induced by  external time dependent
gravitational influences which are what is actually detected.

Let us consider two infinitesimally close geodesics,
$X_{1}^\mu$ and  $X_{2}^\mu=X_{1}^\mu+\eta^\mu$, whose unit tangent vector is $\displaystyle\frac{dx^\alpha}{d\tau}$,
where $\tau$ is their common proper time. The quantity 
$\eta^{\mu}$ is referred to as the connecting vector. 
Theories of detectors start with the equation of geodesic deviation (or Jacobi equation) ~\footnote{For any tensorial quantity $T^{\dots}_{\dots}$, we put 
$\displaystyle\frac{D T^{\dots}_{\dots} }{d\tau} = T^{\dots}_{\dots ;\nu}\frac{dx^\nu}{d \tau}$\,.}, 
\beq 
\frac{D^2\eta^\mu}{d\tau^2} +
R^\mu_{\, \alpha \nu \beta} \frac{d x^\alpha}{d \tau} \frac{dx^\beta}{d \tau}
\eta ^\nu =0 \,,   
\label{deviation}
\eeq 
and then modify it with elastic and damping terms (see, e.g., eqn 4  of \cite{Gibbons:1972}). The connecting vector  satisfies
\beq
g_{\mu \nu} \frac{dx^\mu}{d \tau} \eta^\nu =0.
\eeq

The geodesic deviation has been studied by
Griffiths and Podolsky \cite{GrifPod}. For the central geodesic given by
$U=\tau\,, V=-\half \tau\,,X^1=X^2=0\,,$
their equations may be cast in the form
\begin{subequations}
\begin{align}
 \frac{d^2 \eta^1}{d U^2 }&= 
 \;\;\;\half\cA_+(U) \eta^1 + \half\cA_\times(U)\eta^2\,,
 \\[6pt] 
\frac{d^2 \eta^2}{dU^2 }&= -\half\cA_+(U) \eta^2 +\half\cA_\times(U)\eta^1\,,
 \\[6pt]
\frac{d^2\eta^3}{dU^2}&=0 \,.
\end{align}
\label{GDeviation}
\end{subequations}
 Given $\cA_+(U)$ and $\cA_\times(U)$, this is a system of second order
linear differential equations for $\eta^i$ as a function of $U$, and hence $\tau$. 
 Within  a tubular neighborhood of  any geodesic one may introduce a Fermi coordinate system $(x^0,x^i)$
in which the metric is locally flat and $t=x^0$ coincides with proper time $\tau$ along the geodesic. 
In such a local coordinate system at rest with respect to a freely falling detector, the acceleration of the
separation $\eta^i$ in  such a local coordinate system  frame at rest with respect to is  subject to a forcing term
\beq
-R^i_{\;0j0}\,\eta^j\,,
\eeq
where $0$ labels the time direction and and $i,j$ the spatial directions. 
For a more detailed discussion of Fermi coordinates in plane gravitational
wave spacetimes see \cite{Rakhmanov:2014noa}.

In fact, since,  if $l^i$ is the time averaged separation, 
both the change $x^i=\eta^ i-l^i $ in separation and the curvatures are typically small, one may approximate
the geodesic deviation equation by, 
\beq
\frac{d^2 x^i}{dt^2} =-R^i_{\,0j0}l^j.
\label{d2xi} 
\eeq 
Thus supposing  $\dot{x}^i$ is initially zero, one has an
induced velocity
\beq
v^i(t)= \frac{dx^i}{dt} =
-\int_{t_i}^{t}  dt^\prime R^i_{\;0j0} (t^\prime)l^j\,. 
\eeq

Now in linear theory
\beq
R_{i0j0} = \frac{G}{3 r} \frac{d^4 D_{ij}}{dt^4}(t-r)\,,  \label{source}  
\eeq
where $D_{ij}$ is the quadrupole of the source, $r$ its distance and $u=t-r$ is retarded time. Note that in linear theory and to the approximation we are
using, there is no distinction between upper and lower spatial indices.
Thus for many plausible sources such as
\begin{itemize}
\item gravitational collapse of a previously time independent object to form a black hole
\item or a gravitational flyby
\end{itemize}
the forcing term would be confined to a finite interval 
$t_i \leq t \leq t_f $ 
of time~: it is \emph{pulse-like} referred to as a sandwich wave \cite{Bondi,BoPiRo58,Penr}. It follows that while
the separation $\eta^i$ may have been constant before
the arrival of the pulse, it will  nevertheless, in general, be \emph{time dependent} after the pulse.
In fact it was pointed out in \cite{Gibbons:1972}
that, at the linear level, 
the three  time integrals
of the signal,
\begin{subequations}
\begin{align}
I^{(3)}&=\big(I^{(3)}_{ij}\big)  =\int^{t_f} _{t_i} dt  \int ^{t} _{t_i} dt^\prime  \int^{t^\prime}_{t_i}
dt^{\prime \prime} R_{0i0j}(t^{\prime \prime}) 
\label{I3}
\\
I^{(2)}&=\big(I^{(2)}_{ij}\big)= \int^{t_f}_ {t_i} dt  \int^t_{t_i}   d t^\prime R_{0i0j}(t^\prime) 
\label{I2}
\\
I^{(1)}&=\big(I^{(1)}_{ij}\big) =\int^{t_f}_{t_i} dt   R_{0i0j}(t)
\label{I1}
\end{align}
\label{Iconds}
\end{subequations}
should vanish in the \emph{collapse} case, since $\frac{dD_{ij}}{dt} $ would  vanish initially and finally. 
 By contrast, in the \emph{flyby} case only the last integral needs to vanish,
since initially and finally $D_{ij}$ could be expected to be quadratic in time
and hence only $\frac{dD^3_{ij}}{dt^3}$ would vanish initially and finally.

The analyses of Zel'dovich and Polnarev, and of Braginsky and Grishchuk is entirely at the linear level and as far as the source is concerned,
they simply use the analogue  of (\ref{source}) for the metric perturbation.
Using transverse traceless or radiation gauge \footnote{In fact their  equation (1) is a plane wave in BJR coordinates with $\Tr(a)=2$ which they regard as a
small perturbation of flat space, i.e. when $a_{ij}=\delta_{ij}$.
They write $a_{ij}=\delta _{ij} + h_{ij}$.  } one has
\beq
h^{TT}_{ij} \propto \frac{1}{r} \frac{d^2}{dt^2 }D_{ij}(t-r) \,. \label{metric} 
\eeq
Now from (\ref{metric}) we have
\beq
R_{0i0j} \propto \frac{d^2h^{TT}_{ij}}{dt ^2}\,.
\label{TTrelation}
\eeq  
Thus
\beq
\frac{d^2 x^i}{dt^2} \propto - \frac{d^2 h^{TT}_{ij}} {dt^2} l{}^j 
\eeq
which is consistent with
\beq
x^i \propto  h^{TT}_{ij}l^j \,.
\eeq

Braginsky and Grishchuk also suggest (their equation (7)) that flyby's should have $D_{ij}$ quadratic in time. 
Braginsky and Thorne \cite{BraTho} makes a distinction between two types of bursts, one without memory and one with memory, expressed in terms of a linearised
description of the gravitational perturbation in transverse traceless gauge $h^{TT}_{ij}$ rather than  the gauge-invariant Riemann tensor components $R_{0i0j}$. 
Thus~: 
\begin{itemize}
\item For Gravitational-wave burst without memory $h^{TT}_{ij}$
  is non-zero only in a finite interval $t_i<t<t_f$ 
\item  while for Gravitational-wave burst with memory,
 $h^{TT}_{ij} = {\rm constant}$   for $t>t_f $.
\end{itemize}
From (\ref{TTrelation})
it follows that for bursts \emph{without memory}  the two integrals $I^1$ and $I^2$ in (\ref{Iconds})
should vanish, while for signals \emph{with memory}, only $I^1$ needs to vanish.  
 
To test these ideas we shall consider pulses constructed from Gaussians, and their integrals and derivatives. While not strictly sandwich waves, their curvatures vanish rapidly outside the width of the Gaussian. 
 
\benu

\item 
 For a flyby the $D_{ij}$ could be the third integral of
 a Gaussian and hence $K_{ij}$ would be the \emph{derivative} of a Gaussian, see (\ref{flyby}) below. 
 
\item
 The system considered by Thorne and Braginsky  could  
 be the \emph{second derivative} of a Gaussian, (\ref{Kd2}). 

\item
 For a collapse one could
 take $D_{ij}(u)  \propto -\mathrm{erfc}(u) $, minus the complementary Error Function.
 Thus the Riemann tensor
 or equivalently  $K_{ij}$ would be the \emph{third  derivative} of a Gaussian, (\ref{Kd3}). 
\eenu

%%%%%%%%%%%%%%%%%%%%%%%%%%%%%%%%%%%%%%%%%%%%%%%%%%%%%%%%%%%
\subsection{Memory via Hamilton-Jacobi theory in BJR coordinates}
\label{MemHamJac}
%%%%%%%%%%%%%%%%%%%%%%%%%%%%%%%%%%%%%%%%%%%%%%%%%%%%%%%%%%%

Since it is central to an understanding of the physical reality  of the memory effect,
we shall begin by giving a self-contained account of  the motion
of freely falling particles  using the Hamilton-Jacobi method in BJR coordinates. The results agrees with the derivation in \cite{BlaDam,DGHZ17}, but is included for the sake of making the paper self-contained.
We thus need to solve:
\beq
g^{\mu \nu}\p_\mu S \p_\nu S = 2c \,,
\qquad \frac{dx^\mu}{d \tau}= g^{\mu \nu} \p_\nu  S \,.
\eeq
The coordinates $v, x^i$ are ignorable and we may separate variables  
$$ 
2\,\partial_u S\partial_v S + a^{ij} \p_i S \p_i S =2c,
\qquad
S= G(u) + v p_v + x^i p_i,  
$$ 
where $p_v,p_i$ are constants
 and the motion is reduced to quadratures,
\beq
\frac{dx^i}{d \tau}= a^{ij}p_j \,,\qquad \frac{dv}{d \tau} = \dot G \,,
\qquad \frac{du}{d \tau} = p_v   
\label{HJeqns}
\eeq
where $\tau$ is proper time.
It follows that
{\it  particles which have initially constant 
coordinates  $\bx$, have 
$\bx$ constant for all times for which the coordinates are well-defined}. This is key to our approach to the memory effect. 

In BJR coordinates we may obtain flat spacetime for $u\le u_i$ by
setting $a_{ij}= \delta_{ij}$.
Thus before the pulse arrives we may make this choice. It is
consistent with the Einstein vacuum equations which state that
the trace of the right hand side of  (\ref{Einstein}) should vanish.
However  clearly from  (\ref{Einstein}), {\it we will not have
$a_{ij}=\delta_{ij}$ after the pulse has passed}.

At the linear level
\beq
\ddot{a}_{ij} \approx  2 K_{ij}  \,,
\eeq
and in Brinkmann  coordinates as an exact statement \cite{DGHZ17}   
$ 
K_{ij}=- R_{iUjU} 
$ 
and so at linear level (see (\ref{RBJR}))
we have,
\beq
\ddot a_{ij} {\approx} -2 R_{uiuj} 
\eeq
which is the analogue of (\ref{TTrelation}) in BJR gauge. 
We have
$
\half b_{ij}(u_f) \approx  \displaystyle \int ^{u_f}_{u_i}\!\! du\, K_{ij} (u)  \,. 
$  
Since in linear theory 
$b_{ij} \approx \dot a_{ij}$,
\beq
a_{ij}(u) \, \approx \,
\delta_{ij}  + 2 \int^u _{u_i}\!  du^\prime \int^{u^\prime}_{u_i}\!\!du^{\prime \prime}  K_{ij}(u^{\prime \prime}) \,. 
\label{lintheor}
\eeq

The particles are at rest in this coordinate system; however
their distances apart will be different after the pulse has passed
i.e. for $u>u_f$   \emph{if the metric is different} and hence if the double integral in (\ref{lintheor}) is non-zero.

This, then, is the linear memory effect in BJR coordinates.
A persistent change in the metric means a persistent change in separation. A similar conclusion in the optical context was reached in \cite{Harte15}.

Now it is clear how this works at the non-linear level.
The  equation (\ref{Einstein}) provides  a non-linear
second order differential equation for $a_{ij}$ and with initial conditions
that $a_{ij}= \delta_{ij}$ before the arrival of the pulse. This 
means that in general  $a_{ij} \ne  \delta_{ij}$ after the pulse has passed
and so the distance between nearby freely falling particles
has  altered. At the linear level we can express the shift
in terms of integrals
of the Riemann tensor introduced in \cite{Gibbons:1972}.
In the full non-linear
case  (\ref{Einstein}) has no obvious explicit solution
but in a perturbation expansion, it seems clear that many more such iterated integrals will crop up. In fact, after the first version of this paper was circulated, we were informed that this is indeed the case, see  sec. 4.1.1 of \cite{Harte15}.
 In later sections we shall explore the relevant solutions both numerically and analytically. 

%%%%%%%%%%%%%%%%%%%%%%%%%%%%%%%%%%%%%%%
\section{Geodesics}\label{GeoSec}
%%%%%%%%%%%%%%%%%%%%%%%%%%%%%%%%%%%%%%%

%%%%%%%%%%%%%%%%%%%%%%%%%%%%%%%%%%%%%%%%%%%%%%%%%%%%%%%%%%%%%%%
\subsection{Geodesics in Brinkmann coordinates 
}\label{BrinkmannGeo}
%%%%%%%%%%%%%%%%%%%%%%%%%%%%%%%%%%%%%%%%%%%%%%%%%%%%%%%%%%%%%%%

Brinkmann coordinates, (\ref{Bplanewave}), are convenient
for a numerical study. For simplicity, we only consider the $+$ polarization, for which   
\beq
K_{ij}(U)X^iX^j=
\half\cA_{+}(U)\Big((X^1)^2-(X^2)^2\Big).
\label{Aalone}
\eeq 
The geodesics are solution of the uncoupled system
\begin{subequations}
\begin{align}
&\dfrac {d^2\!X^1}{dU^2} - \half\cA_{+} X^1 = 0,
\label{geoX1}
\\[8pt]
&\dfrac {d^2\! X^2}{dU^2} + \half\cA_{+} X^2 = 0,
\label{geoX2}
\\[8pt]
&\dfrac {d^2\!V}{dU^2} +\frac{1}{4}\dfrac{d\cA_{+}}{dU}\Big((X^1)^2-(X^2)^2\Big) 
+ 
\cA_{+}\Big(X^1\frac{dX^1}{dU}-X^2\dfrac{dX^2}{dU}\Big)=0\,.
\label{geoV}
\end{align}
\label{Bgeoeqn}
\end{subequations}
Fixing the initial conditions $\bX(U_0)=\bX_0$ and $\dot{\bX}(U_0)=\dot{\bX}_0$, the projection of the $4D$  worldline to  the transverse plane is therefore  independent of the choice of $V(U_0) =V_0$, i.e., independent of whether the motion is timelike, lightlike or spacelike. 

The  geodesic deviation equations  of Griffiths and Podolsky's, (\ref{GDeviation}) can be rederived from ours here. For $\eta^i=X_2^i-X_1^i,\, i=1,2$, this follows from the linearity of the first two equations in (\ref{Bgeoeqn}). 
 As to the third one, Eq. (\ref{geoV}) entails that 
\beqa
\frac{d^2(V_2-V_1)}{dU^2}&=&\;
-\frac{1}{4}\frac{d\cA_+}{dU}((X_2^1)^2-(X_1^1)^2-(X_2^2)^2+(X_1^2)^2)
\nn
\\
&&\;\,+\,\cA_+\left(X_2^1\frac{dX_2^1}{dU}-X_1^1\frac{dX_1^1}{dU}-X_2^2\frac{dX_2^2}{dU}-X_1^2\frac{dX_1^2}{dU}\right)
\nn
\\
&=&\;-\frac{1}{4}\frac{d\cA_+}{dU}\left((\eta^1)^2-(\eta^2)^2\right)+\cA_+\Big(\eta^1\dot{\eta}^1-\eta^2\dot{\eta}^2\Big)
\nn
\eeqa
 if one assumes that $X_1^i=0$, i.e., $\eta^i=X_2^i$. The Jacobi deviation equation being linear in $\eta^\mu$, we can conclude that $d^2\eta^3/dU^2=d^2(V_2-V_1)/dU^2=0$. 
 
The system (\ref{Bgeoeqn}) can be solved once $\cA_{+}(U)$ is given. Analytic solutions can be obtained in particular cases only, though, therefore we study our equations numerically. An insight into what happens, is gained by considering Gaussians and their  integrals and derivatives. 
 The colors refer, in all Figs.\ref{Gauss0Geo}-\ref{Gauss1Geo}-\ref{Gauss2Geo}-\ref{Gauss3Geo}, to identical initial conditions   
$\blue{X_0^1}=\red{X_0^2}=\dgreen{V_0}=
\blue{.5},\,\red{1},\,\dgreen{1.5}$ at $U_0<<0$.
 
\begin{itemize}

\item
We start with a toy example, assuming that the gravitational burst is a simple Gaussian,  
\beq
\cA_{+}(U)= \cA^0_{+}(U)\equiv\frac{1}{2}\,e^{-U^2}.
\eeq
Then the integrals (\ref{Iconds}) are
\beq
I^1 = \frac{\sqrt{\pi}}{2}\,{\rm diag}(1,-1), 
\quad
I^2 = I^3  =\infty\,{\rm diag}(1,-1)\, .
\label{Id0}
\eeq
The evolution of the profile and the geodesics are shown
Fig.\ref{Gauss0Profile} and \ref{Gauss0Geo}, respectively.

%%%%%%%%%%%%%%%%%%
\begin{figure}[h]
\begin{center}
\includegraphics[scale=.22]{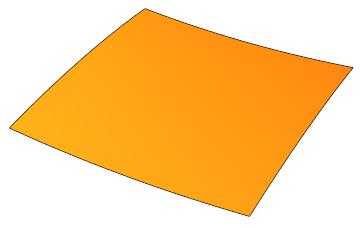}
\includegraphics[scale=.22]{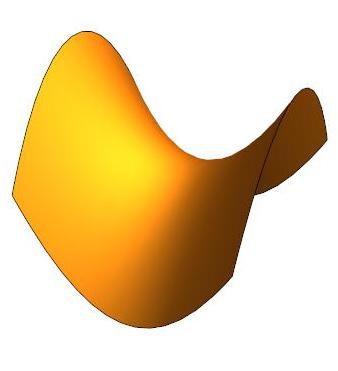}
\includegraphics[scale=.22]{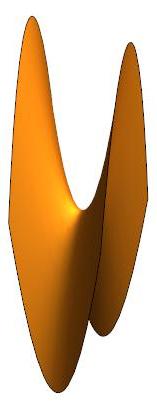}
\includegraphics[scale=.22]{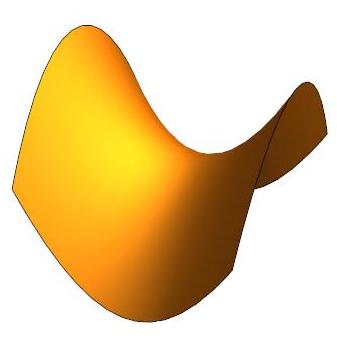}
\includegraphics[scale=.22]{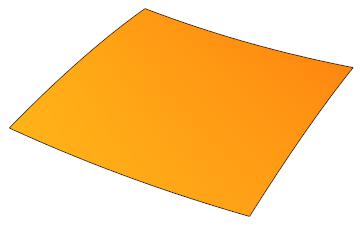}
\end{center}\vskip-8mm
\caption{\textit {Evolution of the wave profile of a Gaussian burst.}}
\label{Gauss0Profile}
\end{figure}
%%%%%%%%%%%%%%%%

%%%%%%%%%%%%%%%%%%
\begin{figure}[h]
\includegraphics[scale=.18]{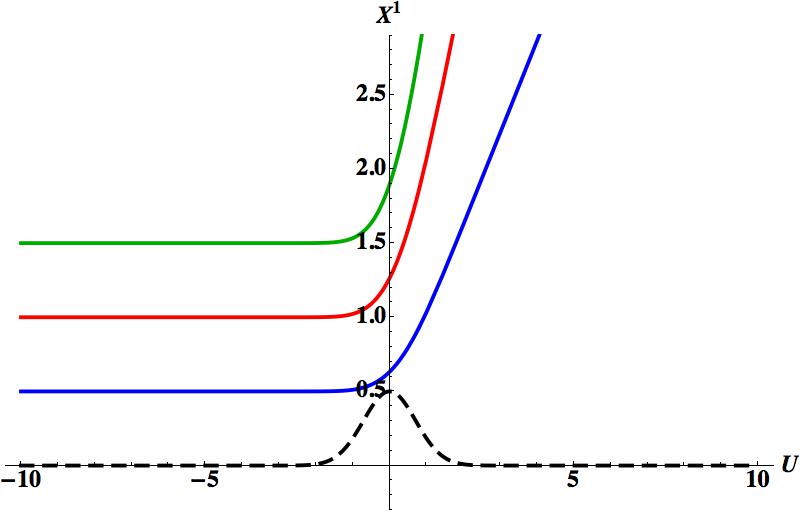} \;
\includegraphics[scale=.18]{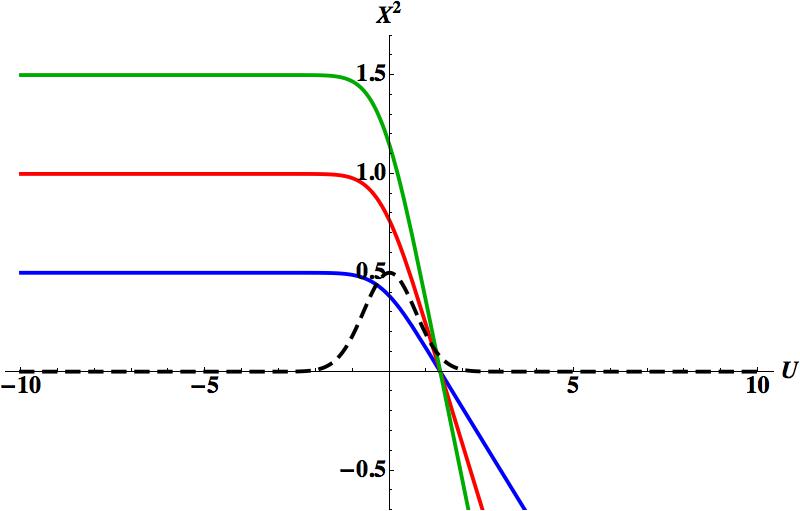} \;
\includegraphics[scale=.18]{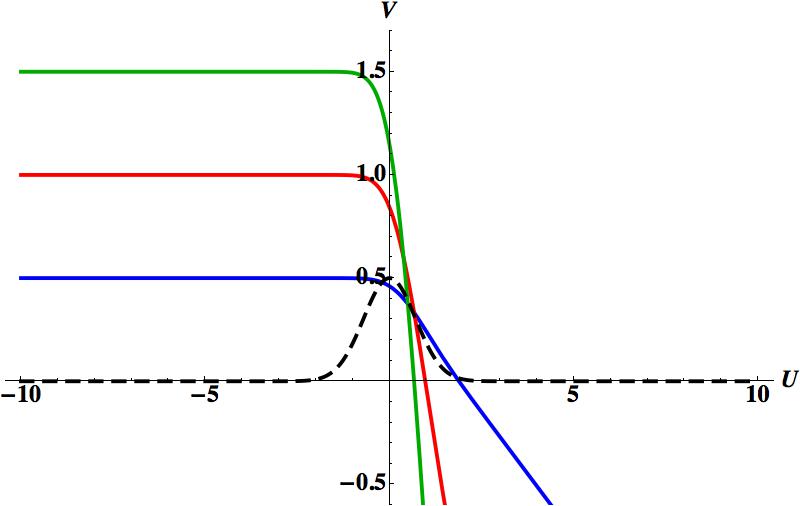}
\\
\caption{\textit{Evolution of  geodesics for a Gaussian burst.
}}
\label{Gauss0Geo}
\end{figure}
%%%%%%%%%%%%%%%%

The variation of the relative (euclidean) distance 
${\Delta}_X(\bX,\bY)= |\bX-\bY| 
$
and of the relative velocity ${\Delta}_{\dot{X}}= |\dot\bX-\dot\bY| $ are
 depicted in Fig.\ref{disB}. The latter could in principle 
 be observed through the Doppler effect \cite{BraGri}.
\begin{figure}[h]
\includegraphics[scale=.16]{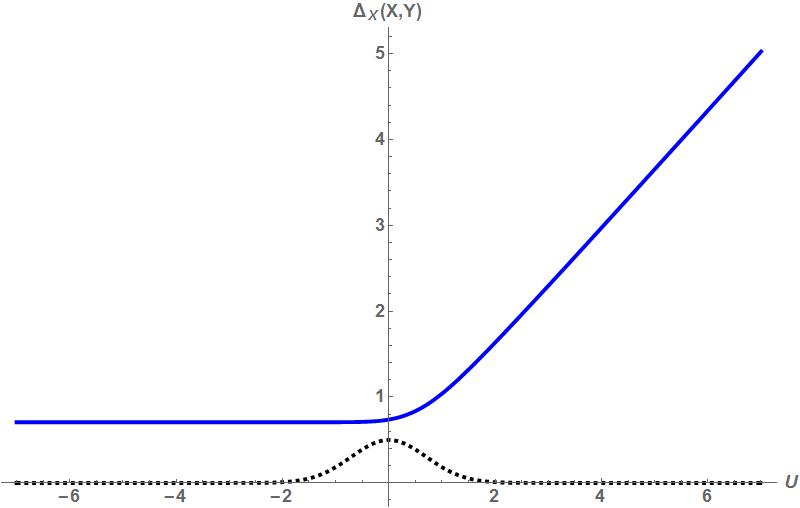}\qquad
\includegraphics[scale=.19]{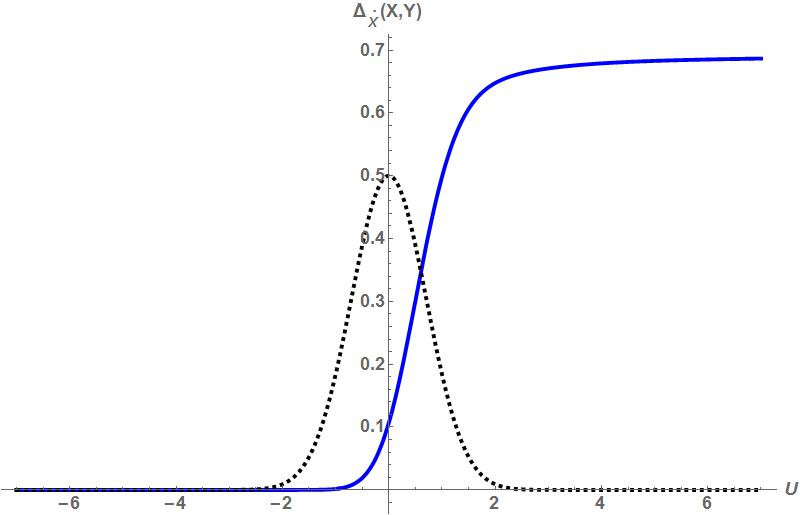}\\
\hskip-3mm
 (a)\hskip 52mm (b)
\vskip-3mm
\caption{\textit{In the Gaussian case, (a) Two particles initially at rest
recede from each other after the wave has passed. Their distance, 
${\Delta}_{X}$, increases roughly linearly in the after-zone.
(b) The relative velocity, ${\Delta}_{\dot{X}}$, jumps to an approximately constant but non-zero value.
}
}
\label{disB}
\end{figure}

\item
For a \emph{flyby} the quadrupole of the source, $D_{ij}$ in (\ref{source}) would be the third integral of
 a Gaussian and hence $\cA_{+}(U)$ would be proportional to the \emph{first derivative} of a Gaussian, 
\beq
\cA_{+}(U) = \cA^1_{+}(U) \equiv
\frac{1}{2}\,\frac{d(e^{-U^2})}{dU}\,. 
\label{flyby}
\eeq
The integrals (\ref{Iconds}) are now
\beq
I^1 = 0,
\quad
I^2 = \frac{\sqrt{\pi}}{2} \,{\rm diag}(1,-1) 
\quad 
I^3 = \infty\,{\rm diag}(1,-1),
\label{Id1}
\eeq
consistently with the interpretation as flyby, cf. sec.
\ref{Detectsec}.
The  geodesics are depicted in FIG.\ref{Gauss1Geo}.
%%%%%%%%%%%%%%%%%%
\begin{figure}[h]
\includegraphics[scale=.18]{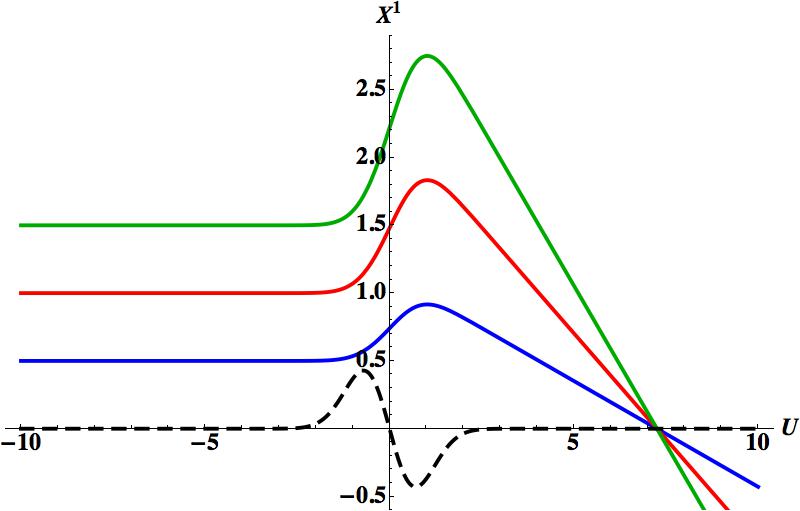}\;
\includegraphics[scale=.18]{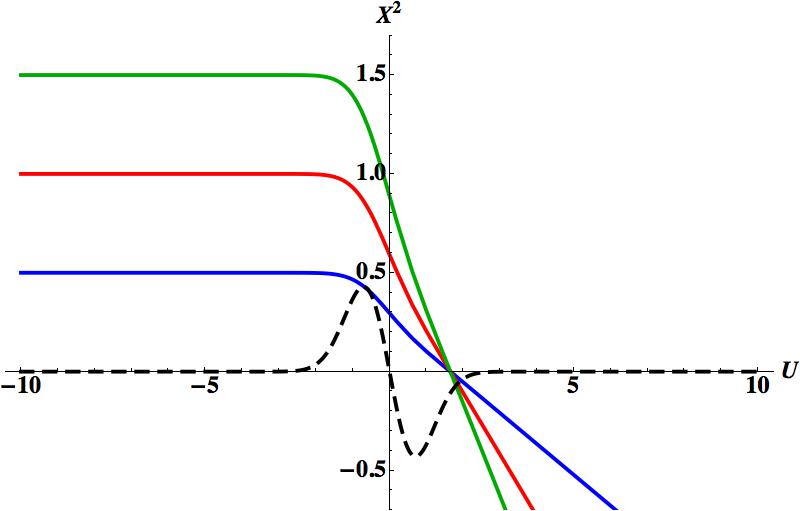}\;
\includegraphics[scale=.18]{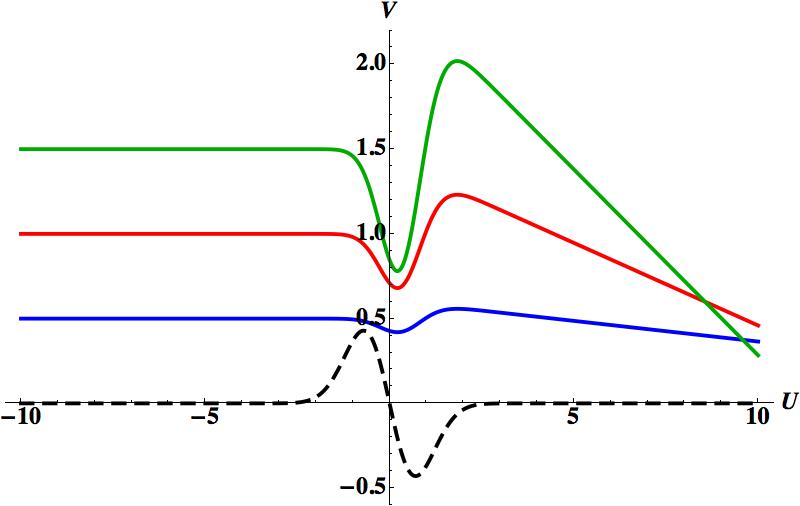}\\
\caption{\textit{Evolution of geodesics for the first derivative of a Gaussian, eqn. (\ref{flyby}), appropriate for flyby.}}
\label{Gauss1Geo}
\end{figure}
%%%%%%%%%%%%%%%%
\goodbreak

\item
 The system considered by Braginsky and Thorne  \cite{BraTho} would seem to correspond to  the \emph{second derivative} of a Gaussian, 
\beq
\cA_{+}(U)= \cA^2_{+}(U)
\equiv \frac{1}{2}\,\frac{d^2(e^{-U^2})}{dU^2}
\,,
\label{Kd2}
\eeq
The integrals (\ref{Iconds}) are now,
\beq
I^1=
I^2 = 0,
\quad
I^3 = \frac{\sqrt{\pi}}{2}\,{\rm diag}(1,-1).
\label{Id2}
\eeq
The geodesics are shown in FIG.\ref{Gauss2Geo}.  
%%%%%%%%%%%%%%%%%%
\begin{figure}[h]
%\hskip-4mm
\includegraphics[scale=.18]{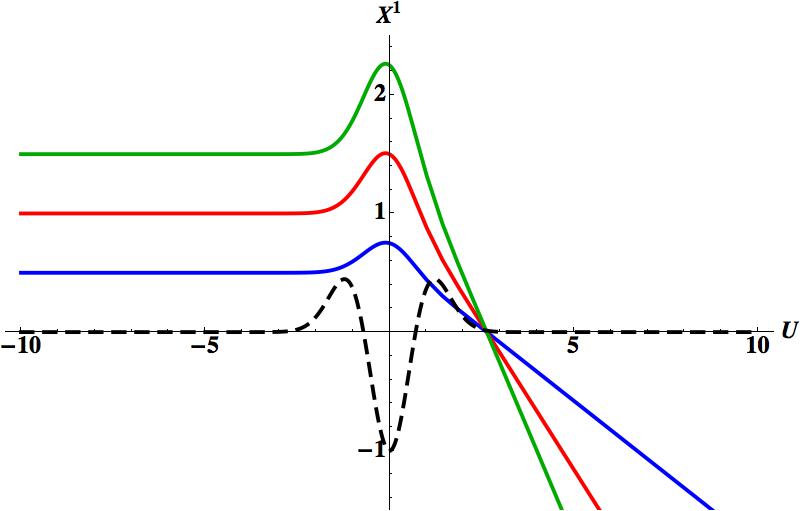}\
\includegraphics[scale=.18]{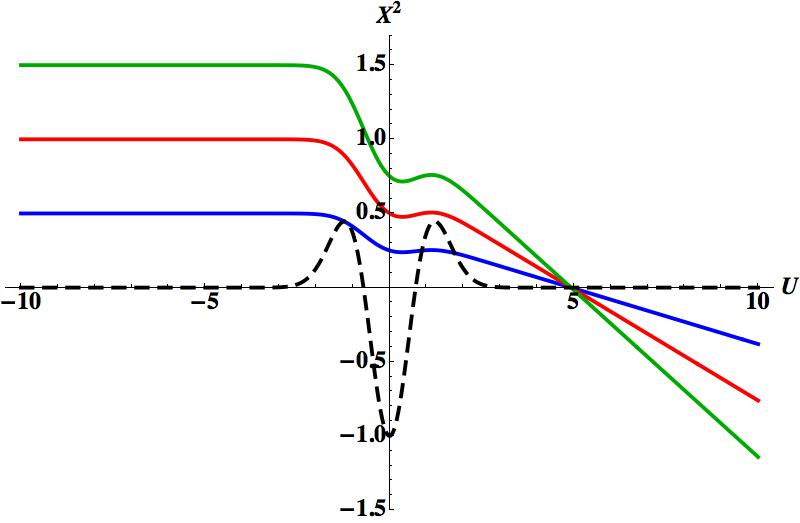}\
\includegraphics[scale=.18]{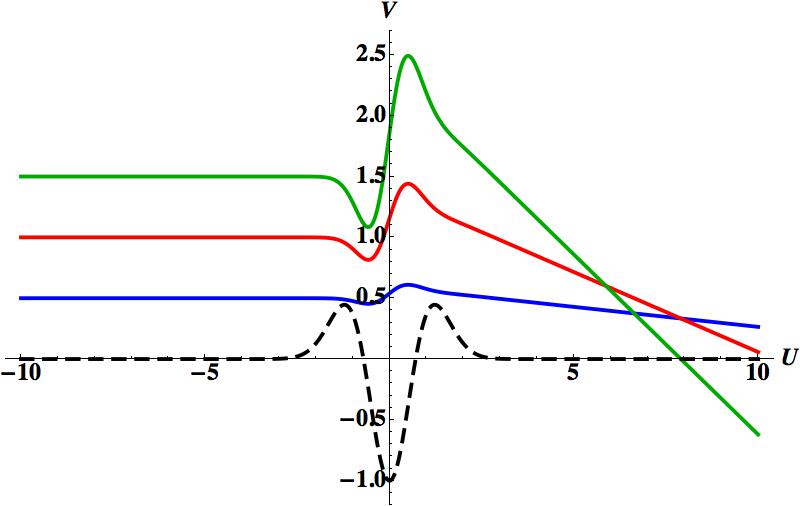}\\
\vskip-5mm
\caption{\textit{The system considered by Thorne and Braginsky  corresponds to the second derivative of a Gaussian, (\ref{Kd2}).}}
\label{Gauss2Geo}
\end{figure}
%%%%%%%%%%%%%%%%
\item
In the early seventies when it was claimed that gravitational wave  bursts had been discovered \cite{Weber} it was suggested  that for gravitational collapse the quadrupole momentum could be modeled by  the fourth derivative of the error function $-{\rm erfc}$ \cite{Gibbons:1972}, yielding,
\beq
\cA_{+}(U) = \cA^3_{+}(U) \equiv
 \frac{1}{2}\,\frac{d^3(e^{-U^2})}{dU^3}\, .
\label{Kd3}
\eeq
%%%%%%
All integrals in (\ref{Iconds})  vanish now,
\beq
I^1 = I^2 = I^3 = 0,
\label{Id3}
\eeq
as expected for gravitational collapse, cf. sec.
\ref{Detectsec}.
The evolution  is  presented  in FIG.\ref{Gauss3Geo}.\vskip-2mm
\begin{figure}[h]
%\hskip-8mm
\includegraphics[scale=.18]{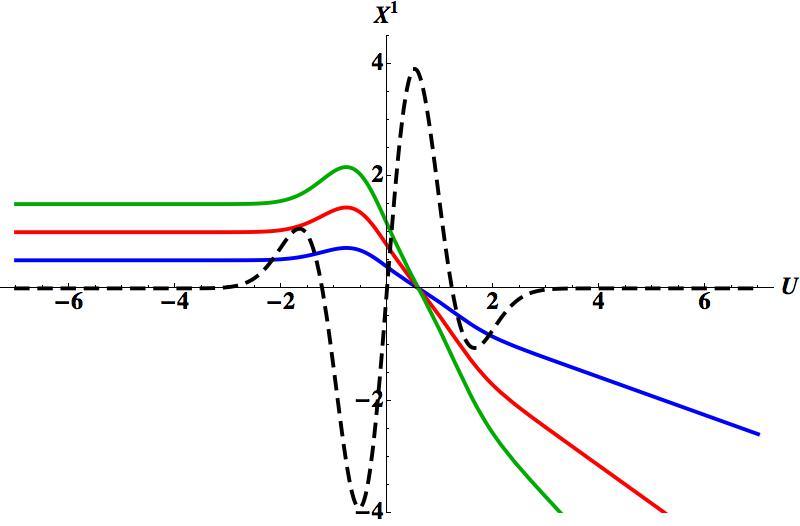}\,
\includegraphics[scale=.18]{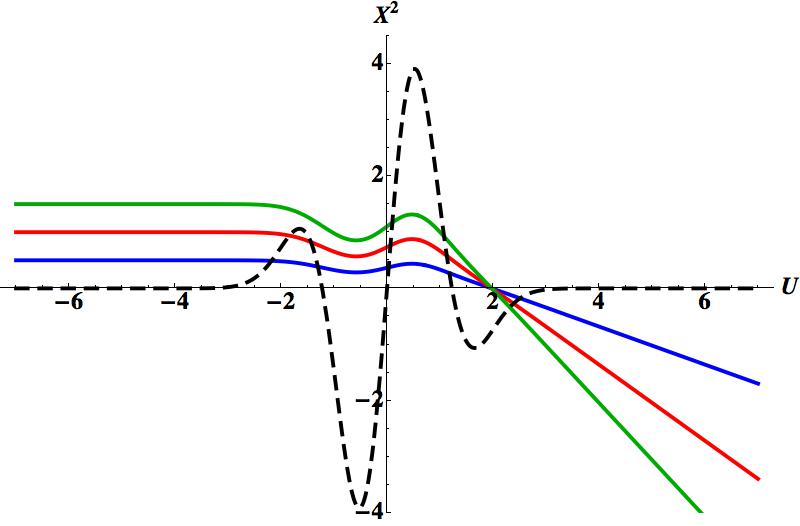}\,
\includegraphics[scale=.18]{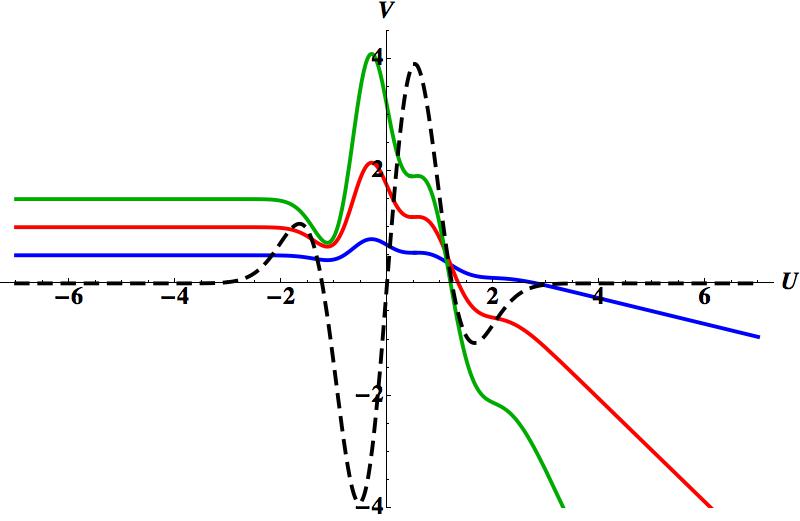}\\
\vskip-5mm
\caption{\textit{Geodesics for particles initially at rest  for $\cA^3_{+}(U)
$\, in (\ref{Kd3}), modelling gravitational collapse. 
}}
\label{Gauss3Geo}
\end{figure}

\end{itemize}

%%%%%%%%%%%%%%%%%%%%%%%%%%%%%%%%%%%%%%%%%%%%%%%%%%%%%%%%%%%%%%%
\subsection{Geodesics in BJR coordinates}\label{BJRGeoSec}
%%%%%%%%%%%%%%%%%%%%%%%%%%%%%%%%%%%%%%%%%%%%%%%%%%%%%%%%%%%%%%%

Further insight can be gained by working in BJR coordinates $(\bx,u,v)$ used in (\ref{BJRmetrics}).

Plane gravitational waves (\ref{Bplanewave}) \blue{or (\ref{BJRmetrics})} have a 5-dimensional isometry group \cite{BoPiRo58}, which has been identified recently
 as the \emph{Carroll group  with broken rotations} \cite{DGHZ17}, implemented on space-time as
\beq\barraynb{lll}
\bx&\to&\bx+H(u)\,\bb+\bc,
\\
u&\to& u,
\\
v&\to& v-\bb\cdot\bx - \2\bb\cdot{}H(u)\,\bb+f,
\earraynb
\label{genCarr}
\eeq
with $\bb,\br\in\bbR^2$ and $f\in\bbR$, where $H(u)$ is the symmetric $2\times 2$ matrix,
\beq
H(u)=\int^u_{u_0}\!\!a(t)^{-1} dt.
\label{Hmatrix}
\eeq
Here $a(u)$ is the transverse-space metric in (\ref{BJRmetrics}) \cite{Sou73}. 

Noether's theorem associates with the Carroll symmetry 5 conserved quantities, associated to these isometries. For the geodesic flow parametrized by some $s$, they are \cite{Sou73,ShortMemory} 
\beq
\bp = a(u)\,\frac{d\bx}{ds},
\qquad
\bk=\bx(u)-H(u)\,\bp,
\qquad
\mu=\frac{du}{ds}\,.
\label{CarCons}
\eeq
An extra constant of the motion we identify with the \emph{kinetic energy}, is
\begin{eqnarray}
\e=\half\rg_{\mu\nu}\frac{dx^\mu}{ds}\frac{dx^\nu}{ds}
=
\half\frac{d\bx}{ds}\cdot a(u(s))\,\frac{d\bx}{ds}
+\mu\frac{dv}{ds}\,.
\label{kinen}
\end{eqnarray}

Geodesics are timelike/lightlike/spacelike, depending on the sign of $\e$. Timelike means $\e<0$, implying that $\mu\neq0$ since $a(u)>0$; the same condition holds also for null geodesics, $\e=0$. Therefore  from now on we put $\mu=1$, 
which amounts to choosing $u$ as parameter. Then the  quantities listed in (\ref{CarCons}) are interpreted as  conserved \emph{linear momentum, boost-momentum and ``$\mu$'' \footnote{When viewed as a Bargmann space of a non-relativistic particle in one lower dimension, $\mu$ (chosen here to be unity) is indeed interpreted as the mass.}.
}

The geodesics may be expressed using the Noetherian quantities above, \cite{Sou73,DGHZ17} via
\begin{equation}
\bx(u)=H(u)\,\bp+\bk,
\qquad
v(u)=-\half \bp\cdot H(u)\,\bp + \e\,u+ d,
\label{BJRGeo}
\end{equation}
where $d$ is a constant of integration.
These equations are consistent with (\ref{HJeqns}) with  $p_v=1$, as expected.
Note that once the values of the conserved quantities are chosen, the only quantity to calculate here is 
 the matrix-valued function $H(u)$ in (\ref{Hmatrix}). Thus the latter
 determines both the action of the isometries and the evolution of causal geodesics.
In flat Minkowski space with the choice $u_0=0$ we have $H(u)=u\,\bone$, yielding free motion 
\beq
\bx(u)=u\,\bp+\bk,
\qquad
v(u)=\left(-\half {\vert\bp\vert}^2  +\e\right) u+ v_0.
\label{Minkmot}
\eeq

\begin{figure}[h]
\includegraphics[scale=.22]{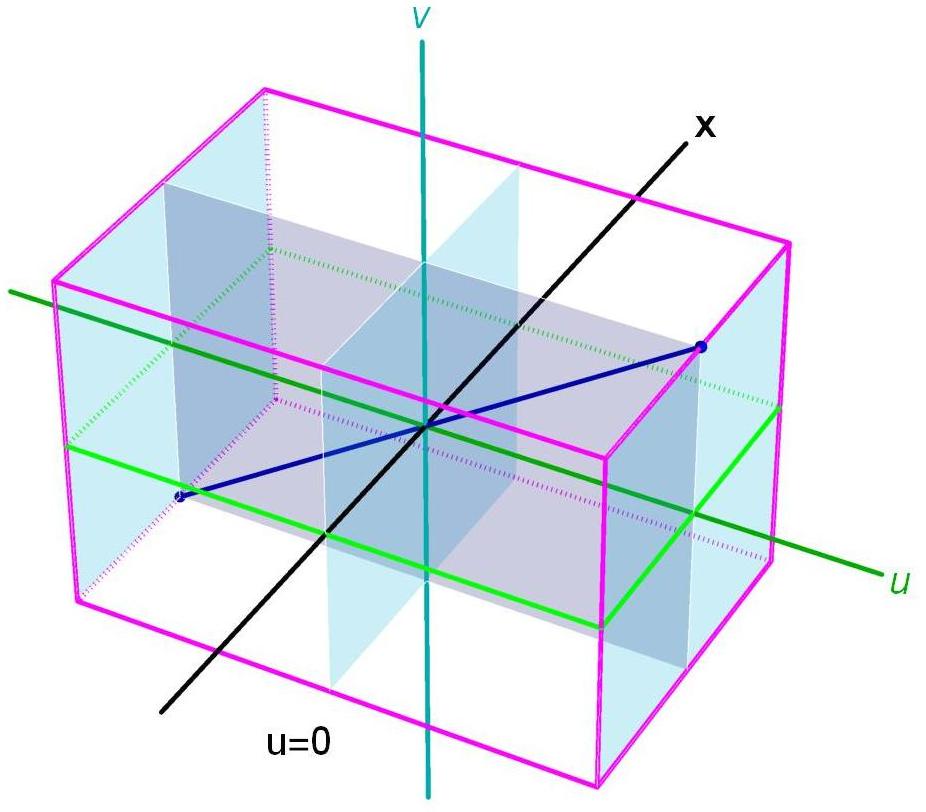}\\
\null\vskip-8mm
\caption{\textit{\small Each geodesic can be ``straightened-out" by a suitable action of the (Carroll) isometry group. 
}}
\label{straightGeo}
\end{figure}

Returning to the general case, the isometries act on the constants of the motion as
 \beq
(\bp,\bk,\e,d) \to (\bp+\bb,\bk+\bc,\e,d+f-\bb\cdot\bk),
\label{Carrolloncq}
\eeq
leaving $\e$ invariant \cite{Sou73,DGHZ17}. 
They can be used therefore to ``straighten out" a geodesic  by carrying it to one with 
$\bp=0,\, \bk=\bx_0$ and $d=0$,
yielding
\beq
\bx(u)=\bx_0=\const,
\qquad v=\e\,u,
\label{fixed}
\eeq 
shown on FIG.\ref{straightGeo}.
Therefore we have, for each sign of $\e$, just one type of ``vertical'' geodesic \cite{Sou73,BoPi89}. 
 Conversely,  any geodesic is obtained from one of form (\ref{fixed}) by an isometry.
\vskip-4mm
%%%%%%%%%%%%%%%%%

%%%%%%%%%%%%%%%%%%%%%%%%%%%%%%%%%%%%%%%%%%%%%%%%%%%%%%%%%%%%%%%%%%%%%%%%%%%
\subsection{The geodesics in the flat before-zone or after-zone}
\label{BefAfterGeo}
%%%%%%%%%%%%%%%%%%%%%%%%%%%%%%%%%%%%%%%%%%%%%%%%%%%%%%%%%%%%%%%%%%%%%%%%%%%%

We first study the geodesics in the flat space-time zones outside a sandwich by making use of the results of Section \ref{PlaneWavesBJRsection}.

Let us suppose that $a(u)=\bone$ in the ``before-zone''  \cite{BoPi89} i.e. for $u<u_i$\,; then the concavity of the function $\chi(u)$ mentioned above implies that the BJR coordinate system suffers a singularity at some time $u_\mathrm{sing}$ such that $\chi(u_\mathrm{sing})=0$, as illustrated on Fig.\ref{chifig}. Note that $u_\mathrm{sing}$ may lie in or outside  the sandwich $[u_i,u_f]$. This coordinate system, used in eqns (\ref{BJRGeo}), is therefore legitimate for $u<u_\mathrm{sing}$ only, which we will assume henceforth.

Consider a system of  particles  at rest (``detectors'', or ``dust'' \cite{Sou73}) in the before-zone. Their geodesics are given, in  natural flat BJR coordinates, by 
\begin{equation}
\hbx=\hbx_0 
\qquad
\&
\qquad
\hv=\e\,(\hu-\,\hu_0)+\hv_0
\label{GeodesicsBeforeZone}
\end{equation}
which identifies the quantities $\hbx_0$ and $\hv_0$ as initial values.

For the flat metric (\ref{aflat}) with general initial condition matrix  $c_0\neq0$, the matrix (\ref{Hmatrix}) is,
\begin{equation}
H(u)=-a_0^{-\half}c_0^{-1}\left[(\bone+(u-u_0)c_0)^{-1}-\bone\right]
a_0^{-\half}\,.
\label{H(u)}
\end{equation}
Then a further tedious calculation yields
the first integrals $\bp,\bk$ and $d$ in  (\ref{BJRGeo}), namely 
\begin{equation}
\bp=-a_0^{\half}c_0\hbx_0,
\qquad
\bk=a_0^{-\half}\hbx_0,
\qquad
d=\hv_0-\e\,\hu_0+\half\hbx_0\cdot{}c_0\hbx_0.
\label{ConstantsOfMotionBeforeZone}
\end{equation} 
%
%%%%%%%%%%%%%%%%%%%%%%%%%%%%%%%%%%%%%%%%%%%%%%%%%%%%%%%%%%%%%%%%%%%%%%%%%%%%
%\subsection{Extending ``flat'' coordinates in the sandwich}
%%%%%%%%%%%%%%%%%%%%%%%%%%%%%%%%%%%%%%%%%%%%%%%%%%%%%%%%%%%%%%%%%%%%%%%%%%%%
%
Moreover, another lengthy calculation yields, using
 (\ref{ConstantsOfMotionBeforeZone}) and (\ref{GeodesicsBeforeZone}), that the geodesics (\ref{BJRGeo}) are expressed, in original BJR coordinates, as 
\vskip-6mm
\begin{subequations}
\begin{align}
\label{xu}
\bx(u)&=\Big[-H(u)\,a_0^{\half}c_0+a_0^{-\half}\Big]\,\hbx,
\\
\label{uu}
u&=\hu,
\\
\label{vu}
v(u)&=\hv+\half\hbx\cdot\Big[c_0-c_0a_0^{\half}H(u)\,a_0^{\half}c_0\Big]\,\hbx.
\end{align}
\label{Newgeoeqn}
\end{subequations}
These equations  may be extended into the sandwich (provided the singularity be avoided) using  $H(u)$  given by (\ref{Hmatrix}). 
Note that  eqns~(\ref{chi-gamma}) and (\ref{oscieqn}) hold everywhere including the inside-zone. 

In the new BJR coordinate system  given by (\ref{Newgeoeqn}) (which we will still denote by~$(\hbx,\hu,\hv)$), the metric can be recast into the form
\begin{subequations}
\begin{align}
\rg&=d\bx\cdot{}a(u)d\bx+2du\,dv
=d\hbx\cdot{}\hat{a}(u)d\hbx+2d\hu\,d\hv,
\label{flathat}
\\[6pt]
\hat{a}(u&)=\Big(a_0^{-\half}-c_0a_0^{\half}H(u)\Big)a(u)\Big(a_0^{-\half}-H(u)a_0^{\half}c_0\Big),
\label{ha(u)}
\end{align}
\label{flatform}
\end{subequations}
cf.  (\ref{aflat}).

We would like to emphasise that the descriptions in Brinkmann and resp. in BJR coordinates are  consistent: numerical calculations  show that pushing forward to B  coordinates a solution constructed in BJR coordinates yields a trajectory which coincides with the one calculated independently in B coordinates, as long as the BJR coordinate system is regular.

%%%%%%%%%%%%%%%%%%%%%%%%%%%%%%%%%%%%%%
\subsection{Tissot Indicatrices and Gravitational Waves}\label{Cartography}
%%%%%%%%%%%%%%%%%%%%%%%%%%%%%%%%%%%%%%

Textbooks providing an account of  the action gravitational waves on a ring of freely falling particles  are often illustrated by a series of time-frames showing how the ring is squashed and stretched as the wave passes over it. See, e.g., \cite{MiThoWh}. 
 This representation has an interesting connection with Tissot's indicatrix  \cite{Tissot,Gott}, which was originally introduced  in  \emph{cartography} to illustrate the distortions brought about  by map projections.

Suppose we have a projection $\phi : S^2 \rightarrow \mathbb{R}^2$
from surface of the earth to a flat sheet
of paper  equipped  with Cartesian coordinates
$x,y$; let $g_{xx}(x,y),\, g_{xy}(x,y)=g_{yx}(x,y),\, g_{yy}(x,y)$
be the components of the push forward to the flat sheet of
paper of the curved metric on the earth's surface.
Tissot's indicatrix at the point $p \in \IS^2 $ with coordinates
$(x_p,y_p)$ is the ellipse 
\beq
g_{xx}(x_p,y_p)x^2 + 2 g_{xy}(x_p,y_p ) xy + g_{yy}(x_p,y_p) y^2 =1 
\eeq
and is the image under $\phi$  of the unit disc in the tangent space of $\IS^2$ \cite{Tissot,Gott}. 

If for some reason the metric  of the surface of the earth varied with time
then so would Tissot's indicatrix:
\beq
g_{xx}(x_p,y_p,t)x^2 + 2 g_{xy}(x_p,y_p,t) xy + g_{yy}(x_p,x_p,t) y^2 =1
\eeq

Returning to gravitational waves, we note that the
two-dimensional sections of the wave fronts at constant time  in
Brinkmann coordinates are given by $U=\const$,
$V=\const$; in Cartesian coordinates $X^i$ carry a flat, time independent Euclidean metric.
These are mapped into two dimensional sections of the wave fronts at constant times
in BJR co\-ordinates $u=\const$, $v=\const$ by
$ 
X^i=P^i_j(u)\,x^j
$ 
 as in (\ref{BBJRtrans}),
which carry a flat time-dependent  Euclidean metric $a_{ij}(u)$ in $x^i$
coordinates. Note that these  two-surfaces do not in general coincide
in spacetime since while $U$ and $u$ are identical, $V$ and $v$ differ.  

The family of timelike geodesics $x^i=\const$  do \emph{not}
have $X^i=\const$ in Brinkmann coordinates. This means that an \emph{initially} (i.e., before the pulse) circular
disc of geodesics in~$X^i$ coordinates,
$ 
\bX\cdot{}\bX \le 1 \,
$ 
for $U < U_i$, projects to a time \emph{independent}
circle in $x^i$ coordinates,    
$
\bx\cdot{}\bx \le 1 \,
$
\emph{for all $u$, i.e., even  during and after the sandwich, $u\geq u_i$},  
 However their inverse image in Brinkmann coordinates is a \emph{time dependent ellipse},
\beq
1=\bx\cdot\bx=
 \bX\cdot(P P^{T})^{-1} \bX .   
\eeq
Since in Brinkmann coordinates the metric is Euclidean, the coordinates represent proper distance measured within that two-surface. 

\begin{figure}[h]
\begin{center}
\includegraphics[scale=.38]{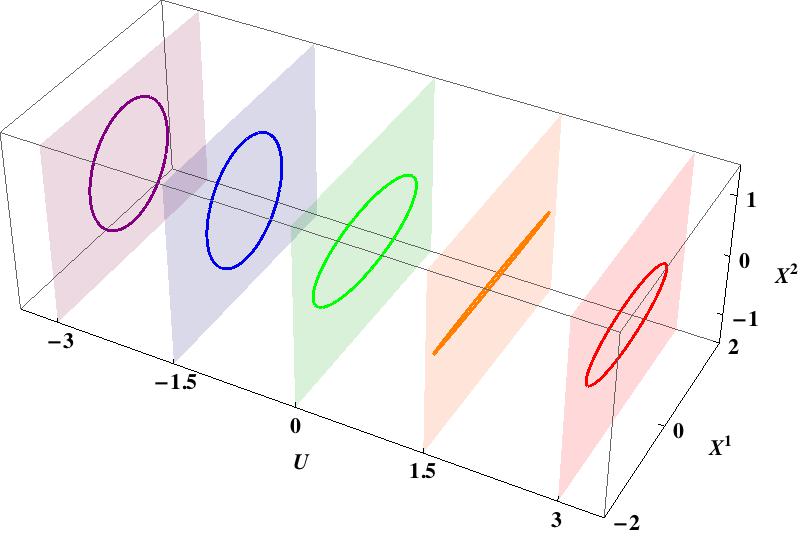} 
\end{center}\vskip-8mm
\caption{\textit{Tissot space-time diagram  for  the linear polarization 
$\cA_+(U) = d(e^{-U^2})/dU,\,\cA_{\times}(U)=0$, 
for  values
$u = -3$ (\textcolor{purple}{purple}), 
 $u = -1.5$  (\blue{blue});
$u = 0$ (\dgreen{green});
 $u = 1.5$ (\textcolor{orange}{orange});
$u = 3$ (\red{red}). }}
\label{Tissotfig}
\end{figure}

The deformation of the Tissot circle is illustrated by the  spacetime diagram in Fig.\ref{Tissotfig} for the linear polarization $\cA_{+}(U) = d(e^{-U^2})/dU,\,\cA_{\times}(U)=0$ appropriate to model flyby, as argued above..
Similar diagrams could be obtained for circular polarization, and also for non burst-like profiles as in the case of primordial gravitational waves.
%%%
The deformation starts before $u=0$, since the burst has a finite thickness. A similar diagram is presented in \cite{ShortMemory} in the gravitation-collapse case 
$\cA_+(U) = d^3(e^{-U^2})/dU^3$.

\subsection{Permanent displacements ?}
%%%%%%%%%%%%%%%%%%%%%%%%%%%%%%%%%%

Eqn. (\ref{BJRGeo}) implies that 
BJR solutions with $\bp=0$ are trivial \emph{for any profile},
\begin{equation}
\bx(u)=\bx_0, 
\qquad
v=\e\,(u-u_0)+v_0
\label{p0geo}
\end{equation}
for all $u$ i.e., in the before, inside and after-zones. This happens in particular for particles which are at rest in the before-zone whose
conserved momentum vanishes because $\bp=a\dot{\bx}$,
cf. (\ref{CarCons}). It is worth emphasizing that the memory effect
\emph{does} arise even in this  case : non-trivial behavior in  B coordinates arises  entirely from the relation \cite{DGHZ17}
\beq
\bX(u)=P(u-u_0)\,\bx_0.
\label{Btraj}
\eeq

But such particles are \emph{not} in general at rest in the
after zone  because $\dot{P}\neq0$ in general -- whereas
some important papers on the memory effect \cite{ZelPol,BraGri,Christo}
 predict  precisely that: particles at rest in the
before zone  could  end up at \emph{rest} but \emph{displaced} in the afterzone.
 Indeed according to some authorities,
  this is taken as a definition of the memory effect.

A possible  indication that this might not be possible
 comes from the  particular cases  studied in sect. \ref{BrinkmannGeo} 
 which show constant but non-zero asymptotic velocity in the
 afterzone (except for $\bx_0=0$).
 Moreover, the relative velocities depend on $\bx_0$, contradicting the expectations of Zel'dovich and Polnarev \cite{ZelPol} cited in the Introduction.

 One may ask whether one may have a smooth  interpolation
  between $P(u)=\bone$ in the before zone and a constant diagonal matrix $P_{\infty}\neq\bone$ in the afterzone, for example, is there a smooth function
 $f(u)$ s.t. 
\beq
P(u)=(1-f(u))\bone+f(u)\,P_{\infty}
\quad\text{with}\quad
f(u)=\left\{\barraynb{ll}
0 &u\leq u_i
\\
1 &u\geq u_f
\earraynb\right. \quad ?
\label{interpol}
\eeq
If we further assume that  $P_{\infty}$ is diagonal, 
$P_{\infty}=\diag(\pi_1,\pi_2)$ with $ \pi_{1,2}=\const\neq 1$ we find that,
\beq
K(u)=\ddot{P}.P^{-1}=
\ddot{f}(u)\,
\diag\Big(
\frac{-1 + \pi_1}{1 - f + \pi_1 f}\;,\: 
\frac{-1 + \pi_2}{1 - f + \pi_2 f}
\Big)
\label{Kf}
\eeq
In order to satisfy  the vacuum Einstein equations
$K$ must be traceless which is however readily seen to contradict the assumption that   $f$ is  \emph{smooth}: $f(u)$ should be  linear with non-zero slope in the inside zone, joined by horizontal lines in the before and after zones and therefore non-differentiable at $u=u_i, u_f$.  

If this rather special example could be generalised, 
  one might conclude  that no static displacement is possible  unless some sort of \emph{impulsive waves} with  non-smooth profiles are considered \cite{Penr2, Sou73, Bini, Steinbauer}.

%%%%%%%%%%%%%%%%%%%%%%%%%%
\section{Null geodesics, light cones and global geometry}\label{PenroseSec}
%%%%%%%%%%%%%%%%%%%%%%%%%%

%%%%%%%%%%%%%%%%%%%%%
\subsection{The memory effect and optics}
%%%%%%%%%%%%%%%%%%%%%

So far we have only considered freely falling particles. However, as remarked in \cite{Harte15}, the memory effect  also influences the motion of light. One way to see this is to recall that Maxwell's equations in a curved vacuum spacetime may be interpreted as flat spacetime electrodynamics in an ``impedance-matched'' medium. Using the results of 
\cite{DGHZ17} we see that in BJR coordinates the permittivity $\epsilon^{ab}$ and 
permeability $\mu^{ab}$ (with $\epsilon^{ab}=\mu^{ab}$) satisfy $\epsilon^{ab}=\mu^{ab}=\delta^{ab}$ before the gravitational wave arrives but after it has passed they are given by
\begin{equation}
  \epsilon^{ij} = \sqrt{\det a(u)}\,\left(a(u)^{-1}\right)^{ij},
  \qquad \epsilon^{33}= \sqrt{\det a(u)}, \qquad \epsilon^{3i}=0,
\end{equation}
and since $ \epsilon^{ij}\neq \delta^{ij}$, the wave has left a memory on the effective optical medium.

%%%%%%%%%%%%%%%%%%%%%
\subsection{Light cones and causality}

In an insightful account of the global geometry of
plane gravitational waves  Penrose \cite{Penr} showed that
in general they are not globally hyperbolic and as a consequence they cannot be isometrically embedded into a higher dimensional
flat space with a just a single
time coordinate \cite{Penr}. Penrose mainly worked in Brinkmann coordinates \cite{Bri} 
although he does allude to the existence of BJR
  coordinates which he ascribes to Rosen \cite{Ros}. 

Penrose  obtains, for a sandwich wave, the formula 
\beq
V= F_{ij}(U) X^iX^j + V_0
\label{PenrV}
\eeq
for the light cone of a point $p=(\bX_0,U_0,V_0)$,
where 
the symmetric matrix $F$ with components  
$ 
F_{ij} \approx (U-U_0)^{-1} \delta_{ij}
$ 
 near $p$  must satisfy 
\beq
\dot{F}+F^2 - K  =0.
\label{dotF}
\eeq

Penrose considers the case when $p$ is located in the flat region before
the pulse arrives. He shows that the metric $F$ becomes singular
within a finite  amount of $u$ time \cite{Sou73}.
This allows him to
obtain his non-global hyperbolicity result. He   points out that this
phenomenon is closely related to the singularity  of
BJR coordinates discussed by \cite{Ros,Bondi,BoPiRo58}.

%%%%%%%%%%%%%%%%%%%%%%%%%%%%%%%%%%%%%%%%%%%%%%%%%%%%
%%%%%%%%%%%%%%%%%%%%%%%%%%%%%%%%%%%%%%%%%%%%%%%%%%%%
%\subsubsection{Light cones in Brinkmann coordinates}
%%%%%%%%%%%%%%%%%%%%%%%%%%%%%%%%%%%%%%%%%%%%%%%%%%%%
%%%%%%%%%%%%%%%%%%%%%%%%%%%%%%%%%%%%%%%%%%%%%%%%%%%%

Penrose's results are readily rederived
by translating our result from BJR to Brinkmann coordinates.
\emph{Null geodesics} are characterized by 
\beq
\e=\half{}\rg_{\mu\nu}\,\dot{x}^\mu\dot{x}^\nu=0.
\eeq 
Special null geodesics, defined by the
vanishing of the linear momentum, $\bp=0$, are
 thus simply
\begin{equation}
\bx(u)=\bx_0,
\qquad
v(u)=v_0.
\label{SpeNullGeod}
\end{equation} 
Moreover, (\ref{BBJRtrans}) gives us the image of the special null geodesics (\ref{SpeNullGeod}), namely
\begin{subequations}
\begin{align}
\bX(U)&=P(U)\,\bx_0,
\label{X(U)}
\\
V(U)&=
v_0-\half \bX\cdot\dot{P}P^{-1}\bX.
\label{V(U)}
\end{align}
\label{}
\end{subequations}
Then the $3$-dimensional light-cone in $\bbR^4$ generated by null geodesics through some point is thus defined by the equation
\begin{equation}
V=v_0-\half \bX\cdot{}F\,\bX,
\label{V(U)bis}
\end{equation}
where
$
F=\dot{P}P^{-1}
$
 satisfies (\ref{dotF})  in view of (\ref{SLB}). Our
equations above thus reproduce (VII.1) and (VII.2) of \cite{Penr}  up to a factor $\half$ and a sign, due to different conventions.

Null geodesics in plane gravitational waves have recently received an extended study in~\cite{Harte12,Harte15}.

%\newpage
%%%%%%%%%%%%%%%%%%%%%%%%%%%%%%%%%%%%%%%%
\section{Exact Einstein-Maxwell plane waves}\label{Einstein-MaxwellSec}
%%%%%%%%%%%%%%%%%%%%%%%%%%%%%%%%%%%%%%%%

Exact Einstein-Maxwell plane waves were first considered in \cite{BaJe} in BJR coordinates. Here we shall follow \cite{Balakin:2000wt}.   For 
 the sake of comparison, we will temporarily adhere
to their signature conventions.
Their  metric in Brinkmann coordinates is 
\beq
\rg = -\delta_{ij}\,dX^i dX^j + dUdV -K(\bX,U) dU^2. 
\eeq
Their  vector potential is taken to  be
\beq
A=A_i(U)dX^i= d(A_i X^i) - X^i A^\prime_i(U) dU \,. \label{vector}
\eeq
In fact we shall find it useful to use the last term on the rhs 
of (\ref{vector}) which differs from $A_i(U)dX^i$ by a gauge transformation. The Maxwell field,
\beq
F = A^\prime _i(U) dU \wedge dX^i \,,
\eeq
solves  $\star\,dF =0$.  
Then the Einstein equation is equivalent to  
\beq
\frac{\p^2 K}{\p X^i \p X^i }= 4G A^\prime_i {A^\prime}_i, 
\label{EMaxeq}
\eeq
where $G$ is Newton's constant and we are using Heaviside
units ($4 \pi \epsilon_0 =1$). We choose the solution
\beq
K(\bX,U) = {\cA_{+}}(U)\big((X^1)^2-(X^2)^2\big)
+2{\cA_{\times}}(U)X^1X^2 +
8G\vert{}\bA^\prime(U)\vert^2\big((X^1)^2 + (X^2)^2\big) 
\label{MaxK}   
\eeq
which merely differs from (\ref{genBrink}) in an additional quadratic  (in Bargmann language, a ``time dependent oscillator'' \cite{DBKP,DGH91,ZHAGK}) term.

The passage to BJR coordinates proceeds 
in a  way similar to the pure gravity case, (\ref{BBJRtrans}).

Note that since the gravitational wave and the electromagnetic wave are essentially independent in Brinkmann coordinates,  
we can specify $\cA_{+}(U)$, $\cA_{\times}(U)$ and $A_i(U)$
independently. There is no {\it graviton-photon} or {\it photon-graviton} conversion, even though the metric has back-reacted to the presence of the electromagnetic field.

This looks very different in the  BJR coordinates, though, in which no simple ``Superposition Principle''  holds. A special case is that one can superpose polarization states in Brinkmann coordinates, but not in a literal  fashion in BJR coordinates
\cite{Poplawski:2011uf,Cropp:2010zk,Cropp:2010xe}.

As pointed out in
\cite{Balakin:2000wt} the coupled Einstein-Maxwell system has five Killing fields,
three of which mutually commute -- in fact, the generators of the isometry group found for a pure plane gravitational wave \cite{DGHZ17} -- namely \emph{the Carroll group} with broken rotations, implemented as in  (\ref{genCarr}).

The proof is straightforward~: everything we developed here and in our previous paper \cite{DGHZ17} goes through unchanged. The metric $a=(a_{ij}(u))$ is related to the wavefront $K=(K_{ij}(u))$ in the usual manner ; the only difference is that the tracelessness of $K$ is replaced by (\ref{EMaxeq}). But this doesn't affect the general form of the metric, cf. (\ref{BJRmetrics}), whose isometries span the  Carroll group in $2+1$ dimensions with broken rotations.

\goodbreak
%%%%%%%%%%%%%%%%%%%%%%%%%%%%%%%%%%%%%%%%
\section{Midi-Superspace quantization of plane gravitational waves}\label{MidiQSec}
%%%%%%%%%%%%%%%%%%%%%%%%%%%%%%%%%%%%%%%%

%%%%%%%%%%%%%%%%%%%%%%%%%%%%%%%%%%%%%%%%
\subsection{Midi-Superspace of plane gravitational waves}\label{MidiSec}
%%%%%%%%%%%%%%%%%%%%%%%%%%%%%%%%%%%%%%%%

We have seen that the \emph{Midi-Superspace}\footnote{
Super-space was a term coined by Wheeler to denote
the configuration space of all Riemannian 3-metrics modulo diffeomorphisms. He thought of it as the natural arena for quantum gravity.  Strictly speaking, when  one quantises, one passes to
the reduced phase space, obtained by taking into account the Hamiltonian and diffeomorphism constraints. This amounts to considering the space of Cauchy data, or equivalently, classical histories, that is, classical solutions of the Einstein equations
 modulo diffeomorphism equivalence. A symmetry reduction (but still with infinite dimensions), is called  a midi-superspace. A symmetry reduction to finite dimensions is called
   a mini-superspace. The reader may consult
\cite{Giulini} for a review.}
 of Ricci flat plane gravitational waves
 is parametrised by the two real functions $\cA_{+}(U)$ and $\cA_{\times}(U)$.
This is an infinite dimensional vector space ${\cal W}$ and in what follows
it will be convenient to assume that $\cA_{+}(U)$ and $\cA_{\times}(U)$ are in
$L^2(\mathbb{R})$ so as to permit Fourier analysis.
Thus we take
\beq
{\cal W}= L^2(\mathbb{R}) \oplus L^2(\mathbb{R})\,.
\eeq
The rotation group $SO(2)$ acts on ${\cal W}$, 
\beq\barraynb{ll}
X^1 &\rightarrow \cos \alpha X^1 + \sin \alpha X^2 
\\
X^2 &\rightarrow \cos \alpha X^2  - \sin \alpha X^1 \,
\earraynb
\Rarrow
\barraynb{ll}
\cA_{+}  &\rightarrow \cos 2\alpha \cA_{+}  - \sin 2\alpha \cA_{\times}
\\
\cA_{\times}  &\rightarrow \cos 2\alpha \cA_{\times}   + \sin 2\alpha \cA_{+} \,
\earraynb\,.
\label{XABrot}
\eeq
Thus ${\cal W}$ carries a helicity $2$ representation of $\mathrm{SO}(2)$, as expected. Note that two metrics related by a rotation are
geometrically identical but we choose to distinguish them because
the action of the rotation does not tend to the identity at infinity.
In other words we are imagining some reference system ``at infinity''
relative to which it is meaningful to speak of the orientation 
$X^1-X^2 $ space.

The real vector space ${\cal W}$ admits a symplectic form $\Omega$.
Let us introduce the notation $\cC=(\cA_{+},\cA_{\times})$ for a general vector
in ${\cal W}$. Then for two vectors $\cC_1$ and $\cC_2$ we define 
\beq
\Omega (\cC_1 ,\cC_2 ) = \int ^\infty_{-\infty} \Bigl (
{\cA_{+}}_1 \frac{d{\cA_{+}}_2 }{dU} -   
{\cA_{+}}_2  \frac{d{\cA_{+}}_1}{dU}+{\cA_{\times}}_1 \frac{d{\cA_{\times}}_2}{dU} -   
{\cA_{\times}}_2\frac{d{\cA_{\times}}_1}{dU}\Bigr) dU.
\eeq

Note that if one regards $V$ as the time coordinate, then ${\cal W}$ is
well defined and independent of $V$ and therefore
the symplectic form $\Omega$ is independent of ``time''.
The hypersurfaces $U=\const$, while not null,
act here as surrogates for Cauchy surfaces.\footnote{ Our choice
  of the sign in front of $2dU dV$  in our metric complicates
  this because it implies that either
  $V$ or $U$ decreases into the future! In order to ensure
  that  $\rg = -dT^2 + dZ^2 +\cdots$ we need to put
  $U=\frac{1}{\sqrt{2}}(Z +T), V=\frac{1}{\sqrt{2}}(Z -T)$  or
  {\it vice versa}, for example. Often $U$ and $V$
  are thought of as retarded and advanced times, i.e. $U=T-Z$
  and $V=T+Z$. This does not quite
  work with our conventions.}

In order to quantize this sector of quantum Einstein theory,
we now pass to the complexification $W_{\mathbb{C}}$ of the classical
real symplectic vector space  ${\cal W}$ and to extend $\Omega $ to  ${\cal W}_{\mathbb{C}}$
in a $\mathbb{C}$-linear fashion. This enables us to endow
${\cal W}_{\mathbb{C}}$ with a sesqui-linear form 
\beq
\langle \cC | \cC \rangle = \frac{i}{2}  \Omega(\bar \cC,\cC)\,,
\eeq 
where $\bar C $ denotes the complex conjugate of $\cC$.  
However $\langle \cC| \cC \rangle$  is not  positive definite.
In order to render
$\Omega (\cC,\cC)$ positive definite, we must restrict
$\langle \cC| \cC \rangle$ to a
$\mathbb{C}$-linear subspace ${\cal H} \subset{\cal W}_{\mathbb{C}} $
on which $\langle \cC| \cC \rangle$ is positive definite on which $\langle {\cA_{+}}| \cA_{\times} \rangle$ is positive definite.

This is conventionally achieved in quantum field theory by restricting to   functions in  ${\cal W}_{\mathbb{C}}$
which are ``positive frequency'' with respect to the coordinate $U$.
If $U$ is chosen to increase to the future, then that means
that ${\cA_{+}}$ and $\cA_{\times}$ only contain Fourier components with
$\omega <0$.  One then has
\beq
{\cal W}_{\mathbb{C}} = \cH \oplus \bar \cH \,.
\eeq

The space of quantum states $\cH$ in this sector
of the entire Hilbert space of Einstein  Quantum Gravity
may be identified with the vacuum Einstein
equations which are analytically continued to complex values of the Brinkmann coordinates $X^1,X^2,U,V$ which are holomorphic
in the lower half $U$-plane.

One might then envisage  an entire free
``one graviton'' Hilbert space by  considering
gravitational waves moving in all possible directions
but not interacting, the continuous  direct sum  
\beq
\int_{{S^2}} \sin \theta d \theta d \phi \cH_{\bf n} \,, 
\eeq
where ${\bf n} \in S^2 $ labels the direction in space of the plane -waves.  
Following the conventional  rules of
 perturbative quantum field theory one might  then pass to
 the free Fock space based on $\cH$. Free correlation  functions
 would then be defined on symmetric products of the complexified
 plane-wave spacetime. The inclusion of interactions then  however
 presents severe difficulties. Moreover, at  the classical level 
 spacetime singularities are encountered when plane waves collide
 \cite{Khan:1971vh,Griffiths:1991zp} .
 
%%%%%%%%%%%%%%%%%%%%%%%%%%%%%%%%%%%%%%%%%%
\subsection{Stokes parameters and the Poincar\'e Sphere}\label{StokesSec}
%%%%%%%%%%%%%%%%%%%%%%%%%%%%%%%%%%%%%%%%%%

The only covariant treatment of this is at the linear level
and notationally rather complicated \cite{AnBr}. A treatment of electromagnetic waves in a pp-wave background is given in~\cite{Hacyan:2012ky}. See also \cite{Duval13}.  Hence we shall follow
the obvious analogy with the electromagnetic case.
We begin by stating our conventions about Fourier transforms.
For a real valued function of $f(U)$ we define its Fourier transform
$\tilde f(\omega)$  by 
\beq
\tilde f(\omega) = \int_{-\infty} ^\infty f(U) e^{i\omega U} dU \,.  
\eeq
The Fourier inversion theorem states that
\beq
f(U) = \frac{1}{2 \pi} \int ^\infty_{-\infty} \tilde f (\omega) e^{-i\omega U} d\omega =  \frac{1}{\pi} \int ^\infty_0 \mathfrak{Re}\left( |\tilde f(\omega)
| \cos \big(\omega t-\psi(\omega)\big)\right) d\omega \,.    
\eeq
Now in the case of a coherent classical electromagnetic wave
the transverse electric field has two real components $E_1(U)$ and $E_2(U)$
with Fourier transforms $\tilde E_1(\omega) $ and $\tilde E_2(\omega)$
and we shall take their gravitational analogues to be ${\cA_{+}}(U)$  and $\cA_{\times}(U)$
with Fourier transforms $\tilde {\cA_{+}}(\omega) $ and $\tilde \cA_{\times} (\omega)$.
From now on we shall work at fixed $\omega$ and suppress it in
most of the formulae which follow. We define the following four real
Stokes parameters \cite{Stokes} which we combine in a Stokes 4-vector $S^\mu$ given by  
\beq 
(S^0,S^1,S^2,S^3 )= 
(|\tilde {\cA_{+}} |^2 + |\tilde \cA_{\times}|^2,\,
|\tilde {\cA_{+}} |^2 - |\tilde \cA_{\times}|^2,\,
 2\,\frak{Re}\,\tilde {\cA_{+}} \overline{\tilde \cA_{\times}}, 
2\,\frak{Im}\,\tilde {\cA_{+}} \overline{\tilde \cA_{\times}}). 
\eeq
It follows that
\beq
-(S^0)^2 + (S^1)^2 + (S^2)^2 + (S^3)^2 = \eta_{\mu \nu}\,S^\mu S^\nu =0 \,.  
\eeq
That is, for a coherent state, the Stokes 4-vector $S^\mu$ is a future directed
null vector passing through the origin of an auxiliary Minkowski spacetime.
For a statistical ensemble of gravitational waves the definition
of the Stokes 4-vector contains a statistical average  or expectation
value denoted by 
$E[\,\cdot\,]$ , thus
\beq
S^\mu = E \Bigl[
(|\tilde {\cA_{+}} |^2 + |\tilde \cA_{\times}|^2,\,
|\tilde {\cA_{+}} |^2 - |\tilde \cA_{\times}|^2,\,
 2\,\frak{Re}\,\tilde {\cA_{+}} \overline{\tilde \cA_{\times}}, 
2\,\frak{Im}\,\tilde {\cA_{+}} \overline{\tilde \cA_{\times}})
\Bigr].
\eeq
It then follows that $S^\mu$ is future directed timelike or null, i.e.
\beq
-(S^0)^2 + (S^1)^2 + (S^2)^2 + (S^3)^2 \le 0\,.
\eeq
It is possible to encode the Stokes 4-vector in a $2\times 2$ hermitian matrix
positive semi-definite coherence matrix $\rho $ which has some analogies
to a density matrix in quantum mechanics. Indeed if the ensemble
is a quantum ensemble this analogy holds fairly closely.

We set
\beq
\rho=E\Bigl 
[\begin{pmatrix}  
|\tilde {\cA_{+}}|^2 , &
 \tilde {\cA_{+}} \overline{\tilde \cA_{\times}}  \cr
  \overline {\tilde {\cA_{+}}}{\tilde \cA_{\times} },   & 
    |{\tilde \cA_{\times}}|^2  
\end{pmatrix}  
\Bigr] 
=
\half \begin{pmatrix} 
S^0 +S^1, &  S^2 +iS^3  
   \cr
   S^2-iS^3 ,   &
   S^0 - S^1   
   \end{pmatrix}.
\eeq

As long as the Stokes 4-vector $S^\mu$ lies inside the future light cone the Hermitian matrix $\rho$ will be positive definite
since ${\rm tr}\ \rho =S^0 > 0$ and $\det \rho  = - \frac{1}{4}\eta _{\mu \nu}  \,S^\mu S^\nu >0 $.  
If the Stokes 4-vector lies on the light cone then $\det \rho =0$. If one introduces the Jones complex valued 2-vector \cite{Jones}
\beq
J= \begin{pmatrix} \tilde {\cA_{+}} \cr \tilde \cA_{\times} \end{pmatrix},  
\eeq
then
\beq
\rho =  E \bigl[ J J^\dagger \bigr ].
\label{rho}
\eeq
In the coherent case, the  Poincar\'e sphere \cite{Poincare}  is obtained  by normalizing the  Jones 2-vector
\beq
J^\dagger J =1 \,,
\eeq
since this implies that $S^0=1$. The spinor geometry behind this construction has recently been described in \cite{1303.4496}.

In the coherent electromagnetic  case  it is customary
  to describe the polarization states by plotting the curve
$(E_1,E_2)= (\frak{Re}\,\tilde E_1 e^{-i\omega t},\frak{Re}\,\tilde E_2 e^{-i\omega t} )$ in the $(X^1,X^2)$ plane. If one  normalizes the Jones 2-vector
\beq
J= \begin{pmatrix} \tilde E_1 e^{-i\omega t} \cr  \tilde E_2 e ^{-i \omega t}
\end{pmatrix}
\eeq
such
that that 
\beq
J^\dagger J = |\tilde E_1 e^{-i\omega t}  |^2 + |\tilde E_2 e^{-i\omega t}|^2 =1\,,
 \label{norm}\eeq one  may introduce
parameters such that  
\beq
J= \begin{pmatrix} \cos\frac{\theta}{2} e^{i(-\omega t  + \delta _1)}  \cr
\sin\frac{\theta}{2} e^{i(-\omega t + \delta _2)}
\end{pmatrix}
\eeq
Now (\ref{norm}) defines a unit three-sphere  in four dimensional
Euclidean space.  As time progresses points on the three sphere
are moved along the orbits of the $U(1)$ action $J \rightarrow e^{-i\omega t}J$.
However  the angle $\theta$ and the relative phase
$\delta=\delta_2-\delta_1, \, -\pi \le \delta \le \pi $   are unchanged. 
As this happens, the electric vector
$( E_1 , E_2 )$  sweeps out an ellipse lying inside 
a rectangle of sides $(\cos \frac{\theta}{2}, \sin \frac{\theta}{2})$ 
whose major axis makes an angle $\half \arctan (\tan \theta 
\cos \delta)$ with the $E_x$ axis. If $\delta >0$.  The
polarization is right handed, if $\delta \le 0$, left handed.
The orbits in $S^3$ are called Hopf fibres and the space of such orbits
is a the  Poincar\'e sphere. Points on the north  and south poles
$\theta = \pm \frac{\pi}{2}$ respectively,
correspond to plane polarised states and points on the equator
$\theta =0$ to circularly polarised states. The remaining states
are elliptically polarised. All but the plane polarised states have a ``handedness''. 

The foregoing theory may  readily  be adapted to
  the complex polarization gravitational wave  amplitudes
 $\tilde {\cA_{+}} e^{-\omega t}$ and
$\tilde \cA_{\times} e^{-\omega t}$. However there is no direct analogue of
 the electric field vector other than the tensor $K_{IJ}$
 and so the image of the
 electric vector executing an ellipse does not seem
 to have a direct analogue. However an  important aspect brought out
 above is that gravitational waves also have a handedness. It is  this
 handedness of primordial gravitational waves
which is  responsible for the generation of the so-called ``B-mode''
 of the electromagnetic waves making up the Cosmic Background Radiation, whose presence is
predicted by theories of inflation \cite{Kamionkowski:1996ks}.

The mechanism
 for the transfer of gravitational wave energy to electromagnetic wave energy
 is the effect of gravitational waves described in the present paper
 on freely  falling electrically 
 charged particles (free electrons in the primordial
 plasma  around  recombination in the case of the CMB) envisaged as an
 abstract possibility in \cite{Bondi-Pirani}
 \footnote{However it should be pointed out that these authors mention neither the CMB nor
     polarization effects.}. The charged particles are necessary because, as we
   noted in the previous section, there is no direct conversion of gravitational
   waves into electromagnetic waves. One might almost claim that
if the B-mode is observed  then the gravitational memory effect will have albeit indirectly, observed.  

The effect of a  polarized monochromatic  gravitational wave
 may be seen by solving the equations of geodesic deviation (\ref{GDeviation})
 assuming
 \beq
\cA_+= \cC_+ \sin(\omega U) \,,\qquad \cA_\times = \cC_\times \sin (\omega U + \phi) 
\,.
\eeq
where the frequency $\omega$, amplitudes $\cC_+ \,, \cC_\times$ and relative
phase $\phi$  are constants.

We conclude by remarking
on the  analogy between the  use of the  Poincar\'e sphere  and   
the way a 2-state system, up to an over all  phase, corresponds  to the
Bloch sphere \cite{Bloch}. However depending upon the spin
or helicity of the states the action of a physical rotation
through an angle $\alpha $  on the spheres will differ. For
spin $\half$ one has $\delta \rightarrow \delta + 2s \alpha$.    
For  quantum systems
there is a notion of Berry or Aharonov-Bohm transport
\cite{Berry:1984jv,Simon:1983mh,Aharonov:1987gg}.
In the case of spin $1$ states this corresponds to
parallel transport on complex projective space $\mathbb{CP}^2$
\cite{Bouchiat:1987vf}.
However in the case of polarised states in  optics this corresponds to
Pancharatnam Transport \cite{Pancharatnam,Berry,Morandi}. 
Pancharatnam's {\it condition of maximum parallelism} 
between two waves with Jones 2-vector's  is
$J^\dagger J^\prime \ge 0$ and in particular that $J^\dagger J^\prime \ge 0$ 
is real.   
If $J$ and $J+dJ$  are two neighboring states we have
by virtual of the normalization condition $J^\dagger J =1$.
\beq
J^\dagger dJ + d J^\dagger J =0\, 
\eeq 
so we define {\it Pancharatnam Parallel Transport} of the phase  by
\beq
J^\dagger dJ  =0= \bar J_1  d J_1  + \bar J_2 d J_2  \,.
\eeq
Now introduce the stereographic coordinate on $S^2$ by
\beq
\zeta= \frac{J_2}{J_1} = \tan \frac{\theta}{2} e^{-2 i \delta} \,. 
\eeq
Pancharatnam's rule for parallel transport reads
\beq
d \ln Z^1 + \frac{ {\bar\zeta}  d \zeta}{1+ |\zeta|^2} =0
\qquad
\text{\small that is} 
\qquad
i (d \tau + d \delta_1)  + 
\frac{ {\bar\zeta}  d \zeta } {1+ |\zeta|^2}= 0 \, 
\eeq
which corresponds to the $U(1)$ connection and curvature,
\beq
A=-i \frac{ {\bar\zeta}  d \zeta}{1+ |\zeta|^2} 
\qquad\text{\small and}\qquad
F=dA =\frac{- i d {\bar \zeta}  \wedge  d \bar \zeta} { (1+|\zeta|^2 )^2 }\,,
\eeq 
respectively
Parallel transport around a simple closed curve $\gamma$
enclosing a domain $D$ produces total holonomy
\beq
\int_D  F  = \half \Omega \,,  
\eeq 
where $\Omega$ is the solid angle subtended by the loop
$\gamma$ at  the centre of the sphere. The factor of $\half$ arises because
$A$ is the spin connection of the metric on $S^2$, and satisfies
the minimal Dirac requirement:
\beq
\int_{S^2 }F  = 2 \pi\,.
\eeq    
The Levi-Civita connection, whose curvature $2F=K$  is the Gauss curvature
is twice as large and its curvature $2F=K$  is the Gauss curvature
(thought of as an $\frak{so}(2)$ 
 valued 2-form) which satisfies  the Gauss-Bonnet condition
\beq
\int_{S^2} 2F = \int _{S^2} K  = 4 \pi \,.  
\eeq

%\newpage

%%%%%%%%%%%%%%%%%%%%%%%%%%%%%%%%%%
%%%%%%%%%%%%%%%%%%%%%%%%%%%%%%%%%%
\section{Conclusion}\label{ConcSec} 
%%%%%%%%%%%%%%%%%%%%%%%%%%%%%%%%%%
%%%%%%%%%%%%%%%%%%%%%%%%%%%%%%%%%%

In this paper we have clarified the physically important
  notions of ``gravitational  memory'' and of ``soft graviton''
  in a simple  and easily calculable  model
  which nevertheless permits a mathematically rigorous treatment
  which captures all the relevant physics. 
  We present  exact solutions of Einstein's equations describing plane gravitational
  waves of arbitrary polarization in the two most useful coordinates.   
We obtained exact expressions for the geodesics in both sets of coordinates. This allowed us  to exhibit the action of a finite duration pulse
  of gravitational radiation on of freely falling particles
  initially at rest in an inertial coordinate system in a portion of flat  Minkowski spacetime to the past of the pulse.
   
%%%%%  
Integrating the geodesic equations in BJR coordinates became possibly due to their
\emph{manifest Carroll symmetry},  (\ref{genCarr}), leading to the conserved quantities  (\ref{CarCons}).

Plane gravitational waves long been known to have a 5-parameter isometry group \cite{BoPiRo58,Sou73}. The generating Killing vectors have, in Brinkmann  coordinates, the components of our $P$ matrix (\ref{SLB}) as coefficients
\cite{BoPi89,Torre}. 
However, being solutions of a Sturm-Liouville equation, these coefficients are not  known in general. 
 
In BJR coordinates the symmetry is manifest and the associated conserved quantities can be calculated by calculating the matrix $H$ in (\ref{Hmatrix}). The price to pay  is that it is now the correspondence B $\Leftrightarrow$ BJR that requires solving a Sturm-Liouville equation~: the difficulty is thus  transferred to the transformation between the two sets of coordinates.

Particles initially at rest have vanishing momentum $\bp=0$ and their trajectory in BJR coordinates is therefore, \emph{for all smooth wave profiles}, the simply straight one in (\ref{fixed}).
Thus after the pulse their transverse positions remain at rest in the non-inertial BJR coordinate system.
  The memory effect is \emph{not lost} however~: it is encoded in the diffeomorphism which we
  calculate explicitly, relating the past inertial coordinates to
  the future non-inertial coordinate system. This diffeomorphism,
  which is in principle constructible from observations using
  gravitational wave detectors, does
  not tend to identity at infinity.
%%%
  
Flat plane wave solutions of Einstein's vacuum equations eqn (\ref{aflat})  in  non-inertial coordinates, are more general than just  Minkowski and
 may be thought of as soft gravitons dressing the initial Minkowski vacuum state.

The extension to Einstein-Maxwell theory is straightforward and a midi-superspace quantization can be given.

%%%%%%%%
After this paper was submitted, we were informed of the related  work of A. Lasenby
\cite{Lasenby}, who arrived, independently, at similar conclusions. He also called our attention at another important reference on the velocity memory effect \cite{GriPol}. Our results here are fully consistent with those of Grishchuk and Polnarev \cite{GriPol}, and also  with those of Bondi and Pirani \cite{BoPi89}, who state that,
\emph{after the wave  have passed detectors originally at rest will move with constant but not zero relative velocity}.
%%%%%%%%

%%%%%%%%%%%%%%%%%%%%%%%%%%%%%%%%%%%%%%%%%%%%%%%%%%%%%%%%%%%%%%%%%%%%%%%%%%%%%%
\goodbreak

\begin{acknowledgments} 
We are grateful to Paul Lasky for instructing us of LIGO's possible future ability to observe the memory effect. 
Anthony Lasenby kindly informed us of his researches on the velocity memory effect \cite{Lasenby}. Correspondence from Abraham Harte is also acknowledged.
GWG would like to thank the 
{\it Laboratoire de Math\'ematiques et de Physique Th\'eorique de l'Universit\'e de Tours}  for hospitality and the  {\it R\'egion Centre} for a \emph{``Le Studium''} research professor\-ship. PH is grateful for hospitality at the \emph{Institute of Modern Physics} of the Chinese Academy of Sciences in Lanzhou. Support by the National Natural Science Foundation of China (Grant No. 11575254) is acknowledged.
\end{acknowledgments}

%%%%%%%%%%%%%%%%%%%%%%%%%%%%%%%%%%%%%%%%%%%%%%%%%%%%%%%%%%%%%%%%%%%%%%%%%%%%%%
%%%%%%%%%%%%%%%%%%%%%%%%%%%%%%%%%%%%%%%%%%%%%%%%%%%%%%%%%%%%%%%%%%%%%%%%%%%%%%
\goodbreak

%%%%%%%%%%%%%%%%%%%%%%%%%%%%%%%%%%%%%%%%%%%%%%%%%%%%%%%%%%%%%%%%%%%%%%%%%%%%%%
%%%%%%%%%%%%%%%%%%%%%%%%%%%%%%%%%%%%%%%%%%%%%%%%%%%%%%%%%%%%%%%%%%%%%%%%%%%%%%

%%%%%%%%%%%%%%%%%%%%%%%%%%%%%%%%%%%%%%%%%%%%%%%%%%%%%%%%%%%%%%%%%%%%%%%%%%%%%%
%%%%%%%%%%%%%%%%%%%%%%%%%%%%%%%%%%%%%%%%%%%%%%%%%%%%%%%%%%%%%%%%%%%%%%%%%%%%%%

\end{document}